\documentclass[letterpaper,twocolumn,10pt]{article}
\usepackage{usenix}

\usepackage{tikz}
\usepackage{amsmath}

\usepackage{xspace}
\usepackage{listings}
\usepackage{pgf}
\usepackage{tikz}
\usetikzlibrary{arrows,automata}
\usetikzlibrary{arrows.meta}
\usetikzlibrary{shadows}
\usetikzlibrary{calc}
\usetikzlibrary{positioning}
\usetikzlibrary{fit}
\usepackage[vlined,ruled,linesnumbered]{algorithm2e}
\usepackage[noend]{algpseudocode}
\usepackage{amssymb}
\usepackage[labelfont=bf,textfont=bf]{caption}
\usepackage{subcaption}
\usepackage{multirow}
\usepackage{hyperref}
\usepackage{pgfplots}
\usepackage{pgfplotstable}
\usepackage{microtype}
\usepackage{balance}
\usepackage{tabularx}
\usepackage{xcolor}
\usepackage{colortbl}
\usepackage{flushend}
\usepackage{enumitem}
\usepackage[norndcorners,customcolors,nofill]{hf-tikz}
\usepackage{xstring}
\usepackage{adjustbox}
\usepackage{tabularx,booktabs}
\usepackage{array}
\usepackage{graphicx}
\usepackage{makecell}
\newcommand\secref[1]{Sect.~\ref{#1}}

\newcommand\figref[1]{Fig.~\ref{#1}}
\newcommand\figsref[1]{Figs.~\ref{#1}}
\newcommand\tabref[1]{Tab.~\ref{#1}}

\newcommand\appref[1]{Appx.~\ref{#1}}

\newcommand\circom{\textsc{Circom}\xspace}
\newcommand\noir{\textsc{Noir}\xspace}

\newcommand\gnark{\textsc{Gnark}\xspace}
\newcommand\circuzz{\textsc{Circuzz}\xspace}
\newcommand\il{\textsc{CircIL}\xspace}
\newcommand\tool{\textsc{Arguzz}\xspace}
\newcommand\riscv{\textsc{RISC-V}\xspace}
\newcommand\risc{\textsc{RISC Zero}\xspace}
\newcommand\zirgen{\textsc{ZirGen}\xspace}
\newcommand\nexus{\textsc{Nexus}\xspace}
\newcommand\jolt{\textsc{Jolt}\xspace}
\newcommand\pico{\textsc{Pico}\xspace}
\newcommand\spOne{\textsc{SP1}\xspace}
\newcommand\openvm{\textsc{OpenVM}\xspace}

\newcommand\todo[1]{\textcolor{red}{#1}}

\newcommand\code[1]{\lstinline[style=basic]{#1}}

\newcommand\bug[1]{
  \IfEqCase{#1}{%
    {1}{\href{https://hackenproof.com/reports/RISCZKVM-25}{#1}}
    {2}{#1}%
    {3}{\href{https://github.com/nexus-xyz/nexus-zkvm/issues/404}{#1}}%
    {4}{\href{https://github.com/nexus-xyz/nexus-zkvm/issues/368}{#1}}%
    {5}{\href{https://github.com/nexus-xyz/nexus-zkvm/issues/413}{#1}}%
    {6}{\href{https://github.com/a16z/jolt/issues/719}{#1}}%
    {7}{\href{https://github.com/a16z/jolt/issues/741}{#1}}%
    {8}{\href{https://github.com/a16z/jolt/issues/741}{#1}}%
    {9}{\href{https://github.com/a16z/jolt/issues/824}{#1}}%
    {10}{\href{https://github.com/a16z/jolt/issues/833}{#1}}%
    {11}{\href{https://github.com/a16z/jolt/issues/892}{#1}}%
  }[\PackageError{bug}{Undefined option to bug: #1}{}]%
}

\newcommand\fix[1]{
  \IfEqCase{#1}{%
    {A}{\href{https://github.com/risc0/zirgen/pull/238}{#1}}%
    {B}{\href{https://github.com/risc0/risc0/pull/3181}{#1}}%
    {C}{\href{https://github.com/risc0/risc0/pull/3015}{#1}}%
    {D}{\href{https://github.com/nexus-xyz/nexus-zkvm/pull/406}{#1}}%
    {E}{\href{https://github.com/nexus-xyz/nexus-zkvm/pull/369}{#1}}%
    {F}{\href{https://github.com/nexus-xyz/nexus-zkvm/pull/417}{#1}}%
    {G}{\href{https://github.com/a16z/jolt/pull/721}{#1}}%
    {H}{\href{https://github.com/a16z/jolt/pull/823}{#1}}%
    {I}{\href{https://github.com/a16z/jolt/pull/790}{#1}}%
    {J}{\href{https://github.com/a16z/jolt/pull/826}{#1}}%
    {K}{\href{https://github.com/a16z/jolt/pull/834}{#1}}%
    {L}{{#1}}%
  }[\PackageError{fix}{Undefined option to fix: #1}{}]%
}
\definecolor{gray}{rgb}{0.5, 0.5, 0.5}
\definecolor{light-gray}{gray}{0.77}
\definecolor{BrickRed}{rgb}{0.8, 0.25, 0.33}
\definecolor{Black}{rgb}{0.0, 0.0, 0.0}
\definecolor{DarkBlue}{rgb}{0.0, 0.0, 0.55}
\definecolor{Crimson}{rgb}{0.86, 0.08, 0.24}
\definecolor{SlateGrey}{rgb}{0.44, 0.5, 0.56}
\definecolor{lightorange}{HTML}{FFB74D}
\definecolor{blue}{rgb}{0.0, 0.0, 1.0}
\definecolor{magenta}{rgb}{0.79, 0.08, 0.48}
\lstdefinestyle{il}{%
  language         = C,%
  alsoletter       = -,%
  morekeywords     = [1]{inputs, outputs, assert},%
  morekeywords     = [2]{},%
  keywordstyle     = [2]\bfseries,%
  morekeywords     = [3]{},%
  keywordstyle     = [3]\bfseries,%
  keywordstyle     = \bfseries,%
  comment          = [l]{//},%
  commentstyle     = \ttfamily\color{Black!60}\small\lst@ifdisplaystyle\small\fi,%
  basicstyle       = \ttfamily\small\lst@ifdisplaystyle\small\fi,%
  emph             = {},%
  emphstyle        = {\color{teal}\bfseries},%
  stringstyle      = \color{BrickRed},%
  columns          = [c]fixed,%
  aboveskip        = 0mm,%
  belowskip        = 2mm,%
  keepspaces       = true,%
  mathescape       = true,%
  escapechar       = @,%
  tabsize          = 2,%
  numbers          = left,%
  numberstyle      = \tiny\color{Black!70},%
  numbersep        = 4pt,%
  stepnumber       = 1,%
  firstnumber      = 1,%
  showstringspaces = false,%
  captionpos       = b,%
  extendedchars    = true,%
  upquote          = true,%
  abovecaptionskip = 0mm,%
  belowcaptionskip = 0mm,%
  xleftmargin      = 3mm,%
  moredelim        = **[is][{\btHL[fill=light-gray]}]{°}{°},
  morecomment      = [s]{/*}{*/}
}

\lstdefinelanguage{Rust}{%
  sensitive%
, morecomment=[l]{//}%
, morecomment=[s]{/*}{*/}%
, moredelim=[s][{\itshape\color[rgb]{0,0,0.75}}]{\#[}{]}%
, morestring=[b]{"}%
, alsodigit={}%
, alsoother={}%
, alsoletter={!}%
, morekeywords={break, continue, else, for, if, in, loop, match, return, while}  % control flow keywords
, morekeywords={as, const, let, move, mut, ref, static}  % in the context of variables
, morekeywords={dyn, enum, fn, impl, Self, self, struct, trait, type, union, use, where}  % in the context of declarations
, morekeywords={crate, extern, mod, pub, super}  % in the context of modularisation
, morekeywords={unsafe}  % markers
, morekeywords={abstract, alignof, become, box, do, final, macro, offsetof, override, priv, proc, pure, sizeof, typeof, unsized, virtual, yield}  % reserved identifiers
, morekeywords=[2]{Add, AddAssign, Any, AsciiExt, AsInner, AsInnerMut, AsMut, AsRawFd, AsRawHandle, AsRawSocket, AsRef, Binary, BitAnd, BitAndAssign, Bitor, BitOr, BitOrAssign, BitXor, BitXorAssign, Borrow, BorrowMut, Boxed, BoxPlace, BufRead, BuildHasher, CastInto, CharExt, Clone, CoerceUnsized, CommandExt, Copy, Debug, DecodableFloat, Default, Deref, DerefMut, DirBuilderExt, DirEntryExt, Display, Div, DivAssign, DoubleEndedIterator, DoubleEndedSearcher, Drop, EnvKey, Eq, Error, ExactSizeIterator, ExitStatusExt, Extend, FileExt, FileTypeExt, Float, Fn, FnBox, FnMut, FnOnce, Freeze, From, FromInner, FromIterator, FromRawFd, FromRawHandle, FromRawSocket, FromStr, FullOps, FusedIterator, Generator, Hash, Hasher, Index, IndexMut, InPlace, Int, Into, IntoCow, IntoInner, IntoIterator, IntoRawFd, IntoRawHandle, IntoRawSocket, IsMinusOne, IsZero, Iterator, JoinHandleExt, LargeInt, LowerExp, LowerHex, MetadataExt, Mul, MulAssign, Neg, Not, Octal, OpenOptionsExt, Ord, OsStrExt, OsStringExt, Packet, PartialEq, PartialOrd, Pattern, PermissionsExt, Place, Placer, Pointer, Product, Put, RangeArgument, RawFloat, Read, Rem, RemAssign, Seek, Shl, ShlAssign, Shr, ShrAssign, Sized, SliceConcatExt, SliceExt, SliceIndex, Stats, Step, StrExt, Sub, SubAssign, Sum, Sync, TDynBenchFn, Terminal, Termination, ToOwned, ToSocketAddrs, ToString, Try, TryFrom, TryInto, UnicodeStr, Unsize, UpperExp, UpperHex, WideInt, Write}
, morekeywords=[2]{Send}  % additional traits
, morekeywords=[3]{bool, char, f32, f64, i8, i16, i32, i64, isize, str, u8, u16, u32, u64, unit, usize, i128, u128}  % primitive types
, morekeywords=[4]{Err, false, None, Ok, Some, true}  % prelude value constructors
, morekeywords=[3]{AccessError, Adddf3, AddI128, AddoI128, AddoU128, ADDRESS, ADDRESS64, addrinfo, ADDRINFOA, AddrParseError, Addsf3, AddU128, advice, aiocb, Alignment, AllocErr, AnonPipe, Answer, Arc, Args, ArgsInnerDebug, ArgsOs, Argument, Arguments, ArgumentV1, Ashldi3, Ashlti3, Ashrdi3, Ashrti3, AssertParamIsClone, AssertParamIsCopy, AssertParamIsEq, AssertUnwindSafe, AtomicBool, AtomicPtr, Attr, auxtype, auxv, BackPlace, BacktraceContext, Barrier, BarrierWaitResult, Bencher, BenchMode, BenchSamples, BinaryHeap, BinaryHeapPlace, blkcnt, blkcnt64, blksize, BOOL, boolean, BOOLEAN, BoolTrie, BorrowError, BorrowMutError, Bound, Box, bpf, BTreeMap, BTreeSet, Bucket, BucketState, Buf, BufReader, BufWriter, Builder, BuildHasherDefault, BY, BYTE, Bytes, CannotReallocInPlace, cc, Cell, Chain, CHAR, CharIndices, CharPredicateSearcher, Chars, CharSearcher, CharsError, CharSliceSearcher, CharTryFromError, Child, ChildPipes, ChildStderr, ChildStdin, ChildStdio, ChildStdout, Chunks, ChunksMut, ciovec, clock, clockid, Cloned, cmsgcred, cmsghdr, CodePoint, Color, ColorConfig, Command, CommandEnv, Component, Components, CONDITION, condvar, Condvar, CONSOLE, CONTEXT, Count, Cow, cpu, CRITICAL, CStr, CString, CStringArray, Cursor, Cycle, CycleIter, daddr, DebugList, DebugMap, DebugSet, DebugStruct, DebugTuple, Decimal, Decoded, DecodeUtf16, DecodeUtf16Error, DecodeUtf8, DefaultEnvKey, DefaultHasher, dev, device, Difference, Digit32, DIR, DirBuilder, dircookie, dirent, dirent64, DirEntry, Discriminant, DISPATCHER, Display, Divdf3, Divdi3, Divmoddi4, Divmodsi4, Divsf3, Divsi3, Divti3, dl, Dl, Dlmalloc, Dns, DnsAnswer, DnsQuery, dqblk, Drain, DrainFilter, Dtor, Duration, DwarfReader, DWORD, DWORDLONG, DynamicLibrary, Edge, EHAction, EHContext, Elf32, Elf64, Empty, EmptyBucket, EncodeUtf16, EncodeWide, Entry, EntryPlace, Enumerate, Env, epoll, errno, Error, ErrorKind, EscapeDebug, EscapeDefault, EscapeUnicode, event, Event, eventrwflags, eventtype, ExactChunks, ExactChunksMut, EXCEPTION, Excess, ExchangeHeapSingleton, exit, exitcode, ExitStatus, Failure, fd, fdflags, fdsflags, fdstat, ff, fflags, File, FILE, FileAttr, filedelta, FileDesc, FilePermissions, filesize, filestat, FILETIME, filetype, FileType, Filter, FilterMap, Fixdfdi, Fixdfsi, Fixdfti, Fixsfdi, Fixsfsi, Fixsfti, Fixunsdfdi, Fixunsdfsi, Fixunsdfti, Fixunssfdi, Fixunssfsi, Fixunssfti, Flag, FlatMap, Floatdidf, FLOATING, Floatsidf, Floatsisf, Floattidf, Floattisf, Floatundidf, Floatunsidf, Floatunsisf, Floatuntidf, Floatuntisf, flock, ForceResult, FormatSpec, Formatted, Formatter, Fp, FpCategory, fpos, fpos64, fpreg, fpregset, FPUControlWord, Frame, FromBytesWithNulError, FromUtf16Error, FromUtf8Error, FrontPlace, fsblkcnt, fsfilcnt, fsflags, fsid, fstore, fsword, FullBucket, FullBucketMut, FullDecoded, Fuse, GapThenFull, GeneratorState, gid, glob, glob64, GlobalDlmalloc, greg, group, GROUP, Guard, GUID, Handle, HANDLE, Handler, HashMap, HashSet, Heap, HINSTANCE, HMODULE, hostent, HRESULT, id, idtype, if, ifaddrs, IMAGEHLP, Immut, in, in6, Incoming, Infallible, Initializer, ino, ino64, inode, input, InsertResult, Inspect, Instant, int16, int32, int64, int8, integer, IntermediateBox, Internal, Intersection, intmax, IntoInnerError, IntoIter, IntoStringError, intptr, InvalidSequence, iovec, ip, IpAddr, ipc, Ipv4Addr, ipv6, Ipv6Addr, Ipv6MulticastScope, Iter, IterMut, itimerspec, itimerval, jail, JoinHandle, JoinPathsError, KDHELP64, kevent, kevent64, key, Key, Keys, KV, l4, LARGE, lastlog, launchpad, Layout, Lazy, lconv, Leaf, LeafOrInternal, Lines, LinesAny, LineWriter, linger, linkcount, LinkedList, load, locale, LocalKey, LocalKeyState, Location, lock, LockResult, loff, LONG, lookup, lookupflags, LookupHost, LPBOOL, LPBY, LPBYTE, LPCSTR, LPCVOID, LPCWSTR, LPDWORD, LPFILETIME, LPHANDLE, LPOVERLAPPED, LPPROCESS, LPPROGRESS, LPSECURITY, LPSTARTUPINFO, LPSTR, LPVOID, LPWCH, LPWIN32, LPWSADATA, LPWSAPROTOCOL, LPWSTR, Lshrdi3, Lshrti3, lwpid, M128A, mach, major, Map, mcontext, Metadata, Metric, MetricMap, mflags, minor, mmsghdr, Moddi3, mode, Modsi3, Modti3, MonitorMsg, MOUNT, mprot, mq, mqd, msflags, msghdr, msginfo, msglen, msgqnum, msqid, Muldf3, Mulodi4, Mulosi4, Muloti4, Mulsf3, Multi3, Mut, Mutex, MutexGuard, MyCollection, n16, NamePadding, NativeLibBoilerplate, nfds, nl, nlink, NodeRef, NoneError, NonNull, NonZero, nthreads, NulError, OccupiedEntry, off, off64, oflags, Once, OnceState, OpenOptions, Option, Options, OptRes, Ordering, OsStr, OsString, Output, OVERLAPPED, Owned, Packet, PanicInfo, Param, ParseBoolError, ParseCharError, ParseError, ParseFloatError, ParseIntError, ParseResult, Part, passwd, Path, PathBuf, PCONDITION, PCONSOLE, Peekable, PeekMut, Permissions, PhantomData, pid, Pipes, PlaceBack, PlaceFront, PLARGE, PoisonError, pollfd, PopResult, port, Position, Powidf2, Powisf2, Prefix, PrefixComponent, PrintFormat, proc, Process, PROCESS, processentry, protoent, PSRWLOCK, pthread, ptr, ptrdiff, PVECTORED, Queue, radvisory, RandomState, Range, RangeFrom, RangeFull, RangeInclusive, RangeMut, RangeTo, RangeToInclusive, RawBucket, RawFd, RawHandle, RawPthread, RawSocket, RawTable, RawVec, Rc, ReadDir, Receiver, recv, RecvError, RecvTimeoutError, ReentrantMutex, ReentrantMutexGuard, Ref, RefCell, RefMut, REPARSE, Repeat, Result, Rev, Reverse, riflags, rights, rlim, rlim64, rlimit, rlimit64, roflags, Root, RSplit, RSplitMut, RSplitN, RSplitNMut, RUNTIME, rusage, RwLock, RWLock, RwLockReadGuard, RwLockWriteGuard, sa, SafeHash, Scan, sched, scope, sdflags, SearchResult, SearchStep, SECURITY, SeekFrom, segment, Select, SelectionResult, sem, sembuf, send, Sender, SendError, servent, sf, Shared, shmatt, shmid, ShortReader, ShouldPanic, Shutdown, siflags, sigaction, SigAction, sigevent, sighandler, siginfo, Sign, signal, signalfd, SignalToken, sigset, sigval, Sink, SipHasher, SipHasher13, SipHasher24, size, SIZE, Skip, SkipWhile, Slice, SmallBoolTrie, sockaddr, SOCKADDR, sockcred, Socket, SOCKET, SocketAddr, SocketAddrV4, SocketAddrV6, socklen, speed, Splice, Split, SplitMut, SplitN, SplitNMut, SplitPaths, SplitWhitespace, spwd, SRWLOCK, ssize, stack, STACKFRAME64, StartResult, STARTUPINFO, stat, Stat, stat64, statfs, statfs64, StaticKey, statvfs, StatVfs, statvfs64, Stderr, StderrLock, StderrTerminal, Stdin, StdinLock, Stdio, StdioPipes, Stdout, StdoutLock, StdoutTerminal, StepBy, String, StripPrefixError, StrSearcher, subclockflags, Subdf3, SubI128, SuboI128, SuboU128, subrwflags, subscription, Subsf3, SubU128, Summary, suseconds, SYMBOL, SYMBOLIC, SymmetricDifference, SyncSender, sysinfo, System, SystemTime, SystemTimeError, Take, TakeWhile, tcb, tcflag, TcpListener, TcpStream, TempDir, TermInfo, TerminfoTerminal, termios, termios2, TestDesc, TestDescAndFn, TestEvent, TestFn, TestName, TestOpts, TestResult, Thread, threadattr, threadentry, ThreadId, tid, time, time64, timespec, TimeSpec, timestamp, timeval, timeval32, timezone, tm, tms, ToLowercase, ToUppercase, TraitObject, TryFromIntError, TryFromSliceError, TryIter, TryLockError, TryLockResult, TryRecvError, TrySendError, TypeId, U64x2, ucontext, ucred, Udivdi3, Udivmoddi4, Udivmodsi4, Udivmodti4, Udivsi3, Udivti3, UdpSocket, uid, UINT, uint16, uint32, uint64, uint8, uintmax, uintptr, ulflags, ULONG, ULONGLONG, Umoddi3, Umodsi3, Umodti3, UnicodeVersion, Union, Unique, UnixDatagram, UnixListener, UnixStream, Unpacked, UnsafeCell, UNWIND, UpgradeResult, useconds, user, userdata, USHORT, Utf16Encoder, Utf8Error, Utf8Lossy, Utf8LossyChunk, Utf8LossyChunksIter, utimbuf, utmp, utmpx, utsname, uuid, VacantEntry, Values, ValuesMut, VarError, Variables, Vars, VarsOs, Vec, VecDeque, vm, Void, WaitTimeoutResult, WaitToken, wchar, WCHAR, Weak, whence, WIN32, WinConsole, Windows, WindowsEnvKey, winsize, WORD, Wrapping, wrlen, WSADATA, WSAPROTOCOL, WSAPROTOCOLCHAIN, Wtf8, Wtf8Buf, Wtf8CodePoints, xsw, xucred, Zip, zx}
, morekeywords=[5]{assert!, assert_eq!, assert_ne!, cfg!, column!, compile_error!, concat!, concat_idents!, debug_assert!, debug_assert_eq!, debug_assert_ne!, env!, eprint!, eprintln!, file!, format!, format_args!, include!, include_bytes!, include_str!, line!, module_path!, option_env!, panic!, print!, println!, select!, stringify!, thread_local!, try!, unimplemented!, unreachable!, vec!, write!, writeln!}  % prelude macros
}%

\lstdefinestyle{basic}{%
  language         = Rust,%
  alsoletter       = -,%
  morekeywords     = [1]{signal,input,output,public,template,component,var,function,return,if,else,for,while,do,log,assert,include,parallel,pragma,circom,custom_templates,defcolumns,defconstraint,let,if,pub,fn,new,type,func,nil},%
  morekeywords     = [2]{\~or!,neq!,eq!,is-not-zero!,vanishes!,if-not-zero,is-not-zero,is-zero,to\_be\_bytes,bytes32\_to\_field,SetString,AssertIsEqual,AssertIsLessOrEqual,SetUint64,Neg,Compiler,FieldBitLen},%
  keywordstyle     = [2]\color{teal}\bfseries,%
  morekeywords     = [3]{},%
  otherkeywords    = {macro_rules!},%
  keywordstyle     = [3]\color{BrickRed}\bfseries,%
  keywordstyle     = \bfseries\color{DarkBlue},%
  comment          = [l]{//},%
  commentstyle     = \ttfamily\color{Black!60}\small\lst@ifdisplaystyle\small\fi,%
  basicstyle       = \ttfamily\small\lst@ifdisplaystyle\small\fi,%
  emph             = {},%
  emphstyle        = {\color{teal}\bfseries},%
  stringstyle      = \color{BrickRed},%
  columns          = [c]fixed,%
  aboveskip        = 0mm,%
  belowskip        = 2mm,%
  keepspaces       = true,%
  mathescape       = true,%
  escapechar       = @,%
  tabsize          = 2,%
  numbers          = left,%
  numberstyle      = \tiny\color{Black!70},%
  numbersep        = 4pt,%
  stepnumber       = 1,%
  firstnumber      = 1,%
  showstringspaces = false,%
  captionpos       = b,%
  extendedchars    = true,%
  upquote          = true,%
  abovecaptionskip = 0mm,%
  belowcaptionskip = 0mm,%
  xleftmargin      = 3mm,%
  moredelim        = **[is][{\btHL[fill=light-gray]}]{°}{°},
  morecomment      = [s]{/*}{*/}
}
\begin{document}
%-------------------------------------------------------------------------------

%don't want date printed
\date{}

% make title bold and 14 pt font (Latex default is non-bold, 16 pt)
\title{\Large \bf \tool: Testing zkVMs for Soundness and Completeness Bugs}

\author{}
%for single author (just remove % characters)
\author{
 {\rm Christoph Hochrainer}\\
 TU Wien\\
 Vienna, Austria\\
 \href{mailto:christoph.hochrainer@tuwien.ac.at}{christoph.hochrainer@tuwien.ac.at}
 \and
 {\rm Valentin W\"ustholz}\\
 Diligence Security\\
 Vienna, Austria\\
 \href{mailto:}{valentin.wustholz@consensys.net}
 \and
 {\rm Maria Christakis}\\
 TU Wien\\
 Vienna, Austria\\
 \href{mailto:}{maria.christakis@tuwien.ac.at}
} % end author

\maketitle

%-------------------------------------------------------------------------------
\begin{abstract}
%-------------------------------------------------------------------------------
  Zero-knowledge virtual machines (zkVMs) are increasingly deployed in decentralized
  applications and blockchain rollups since they enable verifiable off-chain
  computation. These VMs execute general-purpose
  programs, frequently written in Rust, and produce succinct
  cryptographic proofs. However, zkVMs are complex, and bugs in their
  constraint systems or execution logic can cause critical soundness
  (accepting invalid executions) or completeness (rejecting
  valid ones) issues.

  We present \tool, the first automated tool for testing zkVMs for
  soundness and completeness bugs. To detect such bugs, \tool combines
  a novel variant of metamorphic testing with fault injection. In
  particular, it generates semantically equivalent program pairs,
  merges them into a single Rust program with a known output, and runs
  it inside a zkVM. By injecting faults into the VM, \tool mimics
  malicious or buggy provers to uncover overly weak constraints.

  We used \tool to test six real-world zkVMs---\risc, \nexus, \jolt,
  \spOne, \openvm, and \pico---and found eleven bugs in three
  of them. One \risc bug resulted in a \$50,000 bounty, despite prior
  audits, demonstrating the critical need for systematic testing of
  zkVMs.
\end{abstract}

%!TEX root = main.tex

%-------------------------------------------------------------------------------
\section{Introduction}
\label{sec:intro}
%-------------------------------------------------------------------------------

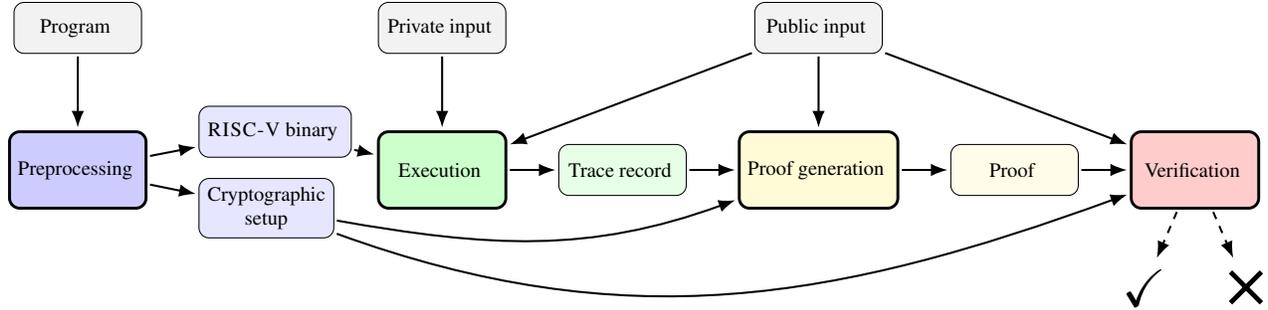
\begin{figure*}[t!]
\centering

\newcommand{\Cross}{\huge$\mathbin{\tikz [x=1.3ex,y=1.3ex,line width=.2ex, black] \draw (0,0) -- (1,1) (0,1) -- (1,0);}$}%
\newcommand{\Checkmark}{\huge$\checkmark$}

\resizebox{0.95\textwidth}{!}{
\begin{tikzpicture}[
  font=\footnotesize,
  % Stage styles with specified colors
  stage/.style={
      draw,
      rounded corners,
      align=center,
      minimum width=6em,
      minimum height=2em,
      fill=#1!20
  },
  component/.style={
      draw,
      rounded corners,
      align=center,
      minimum width=5em,
      minimum height=3em,
      line width=0.4mm,
      fill=#1!20
  },
  % data style (blue boxes)
  data/.style={
    draw,
    rounded corners,
    align=center,
    minimum width=5em,
    minimum height=2em,
    fill=#1!10
  },
  % Control flow arrows (solid black)
  arrowstage/.style={
      -{Triangle[width=8pt,length=6pt]},
      line width=3pt,
      draw=black,
      shorten >=5pt,
      shorten <=3pt
  },
  % Data flow arrows (striked black)
  arrowdata/.style={
      -Latex,
      thick,
      dashed,
      draw=black,
      shorten >=1pt,
      shorten <=1pt
  },
  % relation arrows (striked black)
  arrowcontrol/.style={
      -Latex,
      thick,
      draw=black,
      shorten >=1pt,
      shorten <=1pt
  },
  % relation arrows (striked black)
  arrowrelation/.style={
      -,
      thick,
      dashed,
      draw=black,
  },
  % Host system container
  hostbox/.style={
      draw,
      rounded corners=8pt,
      thick,
      draw=gray!60!black,
      fill=gray!8,
      inner sep=12pt
  }
]

% ---------------------------------------------------------------------------- %
%                                     Setup                                    %
% ---------------------------------------------------------------------------- %

%  --- zkVM Preprocessor --- 
\node[component=blue] (preprocessor) { Preprocessing };

% --- Risc-ish IR ---
\node[data=blue, right=2em of preprocessor, yshift=1.5em] (ir) { \riscv binary };

% --- Cryptographic Parameters ---
\node[data=blue, right=2em of preprocessor, yshift=-1.5em] (crypto) { Cryptographic\\setup };

% ---------------------------------------------------------------------------- %
%                                     Executor                                 %
% ---------------------------------------------------------------------------- %

%  --- zkVM Executor---
\node[component=green, right=9em of preprocessor] (vm) { Execution };

%  --- Trace Record ---
\node[data=green, right=2em of vm] (output) { Trace record };

% ---------------------------------------------------------------------------- %
%                                     Prover                                   %
% ---------------------------------------------------------------------------- %

%  --- Prover---
\node[component=yellow, right=2em of output] (prover) { Proof generation };

%  --- Proof ---
\node[data=yellow, right=2em of prover] (proof) { Proof };

% ---------------------------------------------------------------------------- %
%                                    Verifier                                  %
% ---------------------------------------------------------------------------- %

%  --- Verifier---
\node[component=red, right=2em of proof] (verifier) { Verification };

%  --- Success---
\node[below=2em of verifier, xshift=-2em] (success) {\Checkmark};

%  --- Failure---
\node[below=2em of verifier, xshift=2em] (failure) {\Cross};

% ---------------------------------------------------------------------------- %
%                                  User Input                                  %
% ---------------------------------------------------------------------------- %

% --- User Input / Data ---
\node[data=gray, above=3em of preprocessor] (program) { Program };

% --- Program Private Input ---
\node[data=gray, above=3em of vm] (privinp) { Private input };

% --- Program Public Input ---
\node[data=gray, above=3em of prover] (pubinp) { Public input };

% ---------------------------------------------------------------------------- %
%                                     Edges                                    %
% ---------------------------------------------------------------------------- %

% ------------------------------- preprocessor ------------------------------- %

% --- Program to Preprocessor ---
\draw[arrowcontrol] (program) -- (preprocessor);

% --- zkVM Preprocessor to Cryptographic Parameters ---
\draw[arrowcontrol] (preprocessor) -- (crypto);

% --- zkVM Preprocessor to Risc-ish IR ---
\draw[arrowcontrol] (preprocessor) -- (ir);

% --------------------------------- executor --------------------------------- %

% --- Risc-ish IR to zkVM Executor ---
\draw[arrowcontrol] (ir) -- (vm);

% --- zkVM Executor to Public Output ---
\draw[arrowcontrol] (vm) -- (output);

% --- Private Input to zkVM Executor ---
\draw[arrowcontrol] (privinp) -- (vm);

% --- Public Input to zkVM Executor ---
\draw[arrowcontrol] (pubinp) -- (vm);

% ---------------------------------- prover ---------------------------------- %

% --- Public Input to Prover ---
\draw[arrowcontrol] (pubinp) -- (prover);

% --- Public Output to Prover ---
\draw[arrowcontrol] (output) -- (prover);

% --- Cryptographic Parameters to Prover ---
\draw[arrowcontrol] (crypto) to[out=350,in=200] (prover);

% --- Prover to Proof ---
\draw[arrowcontrol] (prover) -- (proof);

% --------------------------------- verifier --------------------------------- %

% --- Cryptographic Parameters to Verifier ---
\draw[arrowcontrol] (crypto) to[out=340,in=200] (verifier);

% --- Public Input to Verifier ---
\draw[arrowcontrol] (pubinp) -- (verifier);

% --- Proof to Verifier ---
\draw[arrowcontrol] (proof) -- (verifier);

% --- Verifier to Success ---
\draw[arrowdata] (verifier) -- (success);

% --- Verifier to Failure ---
\draw[arrowdata] (verifier) -- (failure);

\end{tikzpicture}}
%\vspace{-1.5em}
\caption{Overview of typical zkVM stages.}
\label{fig:intro-stages}
\end{figure*}

\emph{Zero-knowledge virtual machines} (zkVMs) are emerging as
critical infrastructure for scalable and privacy-preserving
computation, especially in decentralized
applications and blockchain rollups. These VMs enable general-purpose programs to be
executed off-chain while producing succinct, verifiable proofs of
correct execution. zkVMs are complex systems that combine compilers,
execution environments, and cryptographic prover backends---components
that are tightly coupled and heavily optimized for proving performance
and proof size.

More specifically, typical zkVMs execute in four stages shown in
\figref{fig:intro-stages}:
\begin{enumerate}
\item \textbf{Preprocessing:} The program is compiled, often to a
  dialect of the \riscv instruction set. This stage usually also
  includes the cryptographic setup for the subsequent proof generation and
  verification stages.

\item \textbf{Execution:} Given private and public program inputs, the
  execution environment runs the program and records an execution
  trace, called trace record. The trace record contains all
  information necessary to reconstruct the program behavior.

\item \textbf{Proof generation:} Given the public inputs, a
  cryptographic proof is produced based on the trace record and the
  zkVM's constraint system. Note that modern zkVMs use a universal
  constraint system, rather than generating a separate system for each
  program.
  Proving algorithms vary across zkVMs and are often designed to be
  pluggable, allowing support for multiple prover backends with
  different trade offs.
  In addition, many VMs apply proof compression to reduce the size of
  the generated proof, which helps minimize on-chain verification
  costs.

\item \textbf{Verification:} Given the public inputs, the proof is
  checked using the constraint system. If verification succeeds, the
  output is accepted as correct.
  Importantly, verification is decoupled from the proving process and
  can be carried out either by an external verifier or directly
  on-chain through smart contracts, enabling decentralized and
  transparent validation.
\end{enumerate}

Given the complexity and coupling of these stages, bugs in zkVMs are
both likely and difficult to detect. We focus on the two classes of
bugs that developers consider most critical: \emph{soundness} and
\emph{completeness} bugs.
Soundness bugs occur when the zkVM accepts an invalid execution.
These can arise when the constraint system is overly weak; for
instance, the proof may be verified even when the program produces an
incorrect output, thereby compromising the integrity of the
system. Completeness bugs, on the other hand, happen when a valid
execution is incorrectly rejected---often due to overly strict
constraints that rule out legitimate behavior. This degrades user
experience (since user inputs cannot be executed) and may even lead to
liveness issues. Additionally, in some cases, the constraints may
simply be wrong, that is, they may misrepresent the intended semantics
of the VM. Such mismatches can lead to both soundness and completeness
issues, depending on whether they allow invalid executions or reject
valid ones.

Both types of bugs have serious consequences. Soundness bugs can lead
to fraudulent transactions being accepted in blockchain systems,
violating core trust and security assumptions. Completeness bugs can block
legitimate transactions or cause unexpected failures in production. In
both cases, the cost of failure is high, and the complexity of zkVMs
makes these bugs particularly hard to find without thorough testing.

\paragraph{Existing work.}
Recent work~\cite{HochrainerIsychev2025} introduced \circuzz, a fuzzer
for ZK pipelines, such as \circom~\cite{BellesMunozIsabel2023},
\gnark~\cite{Gnark}, and \noir~\cite{Noir}, which compile programs in
domain-specific languages (DSLs) into constraint systems. \circuzz
uses \emph{metamorphic
testing}~\cite{ChenCheung1998,BarrHarman2015,SeguraFraser2016} to
detect soundness and completeness bugs in these pipelines.
The approach starts by generating a deterministic program, called
\emph{circuit}, in \il, an intermediate language designed to express
most features of popular DSLs. Semantics-preserving transformations
are applied to produce a second, equivalent circuit. Both the original
and transformed circuits are then translated into a target DSL,
executed on the same inputs, and their observable behaviors are
compared. Divergences in behavior may indicate overly weak constraints
(soundness bugs), overly strong constraints (completeness bugs), or
issues that can manifest as both soundness and completeness bugs.

zkVMs share many architectural similarities with ZK pipelines but
differ in two key aspects. First, they typically execute
general-purpose (for instance, Rust) programs instead of DSLs. Second,
in zkVMs, developers do not explicitly write constraints using
assertions or DSL primitives that could cause the constraints to
become unsatisfiable, and therefore, prevent proof
generation. Instead, constraints are enforced automatically by the
zkVM based on the semantics of the compiled Rust program, and a proof
is generated. This abstraction improves usability, but also makes it
harder to reason about the enforced constraints and identify bugs.

Adapting the \circuzz approach to zkVMs is desirable, but faces
significant challenges. First, zkVMs are significantly more
computationally expensive, primarily because they must model each
\riscv instruction with precise constraints. Second,
metamorphic testing cannot detect soundness bugs caused by overly weak
constraints when both the original and transformed programs exhibit
the same behavior.
These challenges motivate the need for more efficient and targeted
testing techniques tailored to zkVMs.

\begin{figure*}[t!]
\centering

\newcommand{\Cross}{\huge$\mathbin{\tikz [x=1.3ex,y=1.3ex,line width=.2ex, black] \draw (0,0) -- (1,1) (0,1) -- (1,0);}$}%
\newcommand{\Checkmark}{\huge$\checkmark$}

\resizebox{0.7\textwidth}{!}{
\begin{tikzpicture}[
  font=\footnotesize,
  component/.style={
      draw,
      rounded corners,
      align=center,
      minimum width=8em,
      minimum height=3em,
      line width=0.3mm,
      font=\bfseries\small,
      fill=#1!20
  },
  arrowcomponent/.style={
      -{Triangle[width=8pt,length=6pt]},
      line width=3pt,
      draw=black,
      shorten >=5pt,
      shorten <=3pt
  },
  arrowloop/.style={
      % -Latex,
      -{Triangle[width=8pt,length=6pt]},
      thick,
      line width=2pt,
      dashed,
      draw=black,
      font=\bfseries\small,
      shorten >=5pt,
      shorten <=3pt
  },
  arrowbug/.style={
      -Latex,
      thick,
      dashed,
      draw=black,
      shorten <=3pt
  },
  label/.style={
    circle,
    draw,
    line width=1pt,
    fill=black!60,
    inner sep=0.1em,
    font=\bfseries\small\color{white},
  },
  container/.style={
    draw,
    rounded corners,
    fill=blue!5,
    inner sep=5pt,
  },
  containertitle/.style={
    anchor=south west,
    font=\bfseries\small,
    inner sep=1pt,
  }
]

% ---------------------------------------------------------------------------- %
%                               CircIL Container                               %
% ---------------------------------------------------------------------------- %

\node[container,
  minimum width=21.4em,
  minimum height=5.5em,
  anchor=north west,
  xshift=-5em,
  yshift=3.3em
] (circil) {};
\node[containertitle, yshift=-1em] at (circil.north west) {\il};

% ---------------------------------------------------------------------------- %
%                              Circuit Generation                              %
% ---------------------------------------------------------------------------- %

%  --- Circuit Generation ---
\node[component=gray] (circgen) { Circuit generation };
\node[label, anchor=south east, xshift=0.5em, yshift=-0.5em] at (circgen.north west) {1};

% ---------------------------------------------------------------------------- %
%                            Circuit Transformation                            %
% ---------------------------------------------------------------------------- %

%  --- Circuit Transformation ---
\node[component=gray, right=2em of circgen] (circtransform) { Circuit transformation };
\node[label, anchor=south east, xshift=0.5em, yshift=-0.5em] at (circtransform.north west) {2};

% ---------------------------------------------------------------------------- %
%                              Circuit Translation                             %
% ---------------------------------------------------------------------------- %

%  --- Circuit Translation ---
\node[component=gray, right=2em of circtransform] (circtranslate) { Circuit-to-Rust\\translation };
\node[label, anchor=south east, xshift=0.5em, yshift=-0.5em] at (circtranslate.north west) {3};

% ---------------------------------------------------------------------------- %
%                          Product-Program Generation                          %
% ---------------------------------------------------------------------------- %

%  --- Product-Program Generation ---
\node[component=gray, right=2em of circtranslate] (prodgen) { Product-program \\ generation };
\node[label, anchor=south east, xshift=0.5em, yshift=-0.5em] at (prodgen.north west) {4};

% ---------------------------------------------------------------------------- %
%                               Input Generation                               %
% ---------------------------------------------------------------------------- %

%  --- Input Generation ---
\node[component=gray, below=5em of prodgen] (inpgen) { Input generation };
\node[label, anchor=south east, xshift=0.5em, yshift=-0.5em] at (inpgen.north west) {5};

% ---------------------------------------------------------------------------- %
%                      Normal Execution & Trace Collection                     %
% ---------------------------------------------------------------------------- %

%  --- Normal Execution & Trace Collection ---
\node[component=gray, left=2em of inpgen] (exec) { Normal VM execution\\and trace collection };
\node[label, anchor=south east, xshift=0.5em, yshift=-0.5em] at (exec.north west) {6};

% ---------------------------------------------------------------------------- %
%                                Weird Execution                               %
% ---------------------------------------------------------------------------- %

%  --- Weird Execution ---
\node[component=gray, left=2em of exec] (weirdexec) { VM execution\\with malicious prover };
\node[label, anchor=south east, xshift=0.5em, yshift=-0.5em] at (weirdexec.north west) {7};

% ---------------------------------------------------------------------------- %
%                                      Bug                                     %
% ---------------------------------------------------------------------------- %

\node[below=5em of circgen] (bug) {\includegraphics[width=30pt]{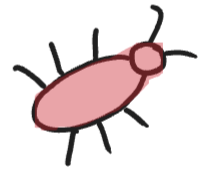}};

% ---------------------------------------------------------------------------- %
%                                     Edges                                    %
% ---------------------------------------------------------------------------- %

% -------------------------------- Main Edges -------------------------------- %

% --- Circuit Generation to Circuit Transformation ---
\draw[arrowcomponent] (circgen) -- (circtransform);

% --- Circuit Transformation to Circuit Translation ---
\draw[arrowcomponent] (circtransform) -- (circtranslate);

% --- Circuit Translation to Product-Program Generation ---
\draw[arrowcomponent] (circtranslate) -- (prodgen);

% --- Product-Program Generation to Input Generation ---
\draw[arrowcomponent] (prodgen) -- (inpgen);

% --- Input Generation to Normal Execution ---
\draw[arrowcomponent] (inpgen) -- (exec);

% --- Normal Execution to Fault Injection ---
\draw[arrowcomponent] (exec) -- (weirdexec);

% -------------------------------- Loop Edges -------------------------------- %

% --- Product-Program Generation to Circuit Transformation ---
\draw[arrowloop] (prodgen.north) -- ++ (0, 2.5em) node[above=0.6em, xshift=-10em] {Next transformation} -| (circtransform);

% --- Product-Program Generation to Circuit Transformation ---
\draw[arrowloop] (weirdexec.north) -- ++ (0, 2em) node[above=0.3em, xshift=10em] {Next input} -| (inpgen.140);

% --------------------------------- Bug Edges -------------------------------- %

% --- Normal Execution to MT Bug ---
\draw[arrowbug] (exec.south) -- ++ (0, -1.5em) -| (bug.south);

% --- Weird Execution to Injection Bug ---
\draw[arrowbug] (weirdexec) -- (bug);

% ---------------------------------------------------------------------------- %

\end{tikzpicture}}

\caption{Overview of \tool.}
\label{fig:overview-arguzz}
\end{figure*}

\paragraph{Our approach.}
In this paper, we tackle these challenges with a new approach for
testing zkVMs that \emph{integrates a novel variant of metamorphic
testing with fault injection}. We implement our approach in a fuzzer
called \tool. To the best of our knowledge, \tool is the first fuzzer
to target zkVMs and to combine these two testing techniques.

On a high level, we design \tool as follows.
First, we adapt the \circuzz framework to zkVMs by translating
circuits generated in \il into semantically equivalent Rust programs
that can run inside zkVMs. Since \il does not support the full
instruction set used by zkVMs, we extend it with custom functions that
include inline assembly. This enables our circuit generator to use all
instructions supported by the target VMs.

Second, to improve efficiency, we introduce a novel variant of
metamorphic testing. Instead of running the original and transformed
programs separately, we merge them into a single Rust program that
compares their results and computes a known, expected output. This
merged program is then executed inside the zkVM. Running a single
combined program increases test throughput and simplifies result
checking, while preserving the ability to detect behavioral
inconsistencies.

Third, to detect soundness bugs due to overly weak constraints---when
both the original and transformed programs exhibit the same
behavior---we develop a fault-injection
mechanism~\cite{ArlatAguera1990,ClarkPradhan1995,HsuehTsai1997}. We
adopt a view where the zkVM consists of two parties: the prover and
the verifier. The prover corresponds to the first three stages
described earlier---preprocessing, execution, and proof
generation---while the verifier is the final stage that checks the
proof (see \figref{fig:intro-stages}). Given this view, we inject
faults into the prover's execution logic to mimic a malicious or buggy
prover---we refer to it simply as malicious in the rest of the
paper. We then execute the same merged Rust program using
this malicious prover and check whether the unmodified verifier
accepts the resulting proof. If the program output diverges from the
expected one (meaning that the fault injection successfully affected
the program execution), but the verifier is deceived into accepting
the proof, this indicates a soundness bug due to underconstrained
behavior.

Overall, \tool identifies three classes of bugs:
\begin{itemize}
\item The merged program crashes during execution on the original
  VM, indicating a completeness bug;
\item The merged program successfully completes on the original VM
  but produces an unexpected output, indicating a soundness bug, a
  completeness bug, or an issue that can manifest as both;
\item The merged program is run on a malicious prover, it produces an
  unexpected output, and the verifier is deceived into accepting the
  proof, indicating a soundness bug.
\end{itemize}

We used \tool to test six popular zkVMs based on the \riscv
instruction set. As we discuss in our experimental evaluation, we
found soundness and completeness bugs across three of these systems,
namely, \risc~\cite{Risc0}, \nexus~\cite{Nexus}, and
\jolt~\cite{ArunSetty2023,Jolt}, for a total of \emph{three soundness and
eight completeness bugs}. One of the soundness bugs we
uncovered in \risc was so severe that it earned a \emph{\$50,000
bounty} from the developers. This is significant because \risc is
already deployed in production and has undergone prior security
audits.

Note that we followed a responsible disclosure process for all
detected issues, either privately reporting them to the development
teams or obtaining permission to disclose them publicly.

\paragraph{Contributions.}
Our main contributions are:
\begin{itemize}
\item A novel testing technique for detecting soundness and
  completeness bugs in zkVMs. Our technique combines an efficient
  variant of metamorphic testing with a fault-injection mechanism that
  mimics a malicious prover.

\item An implementation of this technique in \tool. \tool is designed
  to be modular and can be adapted to new \riscv-based zkVMs with
  modest effort.

\item A practical evaluation across six real-world zkVMs. \tool found
  eleven correctness bugs in total---three soundness and
  eight completeness bugs. One of the \risc soundness bugs
  resulted in a \$50,000 bounty given its critical severity.
\end{itemize}

%% \paragraph{Outline.}
%% %
%% The rest of the paper is organized as follows. In the following
%% section, we give an overview of \tool, and in \secref{sec:approach},
%% we describe the technical details of our
%% approach. \secref{sec:evaluation} presents our experimental
%% evaluation. We discuss related work in \secref{sec:related} and
%% conclude in \secref{sec:conclusion}.

%%% Local Variables:
%%% mode: latex
%%% TeX-master: "main"
%%% End:

%!TEX root = main.tex

%-------------------------------------------------------------------------------
\section{Overview}
\label{sec:overview}
%-------------------------------------------------------------------------------

\begin{figure*}[t!]
\begin{subfigure}[b]{0.525\textwidth}
\begin{lstlisting}[style=il]
inputs : a, b, c @\label{l:cins}@
outputs: out @\label{l:couts}@
out = (a % (b + c)) @\label{l:cout}@
\end{lstlisting}
\caption{Circuit $C_1$ in \il.}
\vspace{1em}
\label{fig:ex-il1}
\end{subfigure}%
\begin{subfigure}[b]{0.475\textwidth}
\begin{lstlisting}[style=il]
inputs : a, b, c
outputs: out
out = (a % ((c + 0) + b))
\end{lstlisting}
\caption{Circuit $C_2$ in \il.}
\vspace{1em}
\label{fig:ex-il2}
\end{subfigure}
\begin{subfigure}[b]{0.525\textwidth}
\begin{lstlisting}[style=basic]
fn c1(a: u32, b: u32, c: u32) -> u32 {
  a % (b + c)
}
\end{lstlisting}
\caption{Circuit $C_1$ as a Rust function.}
\label{fig:ex-rust1}
\end{subfigure}%
\begin{subfigure}[b]{0.475\textwidth}
\begin{lstlisting}[style=basic]
fn c2(a: u32, b: u32, c: u32) -> u32 {
  (a % ((c + 0) + b))
}
\end{lstlisting}
\caption{Circuit $C_2$ as a Rust function.}
\label{fig:ex-rust2}
\end{subfigure}
\caption{Example \il circuits and Rust functions generated by \tool.}
\label{fig:example}
\end{figure*}

Our fuzzer tests zkVMs by combining an efficient variant of
metamorphic testing and fault injection. \tool proceeds through the
following seven steps, summarized in \figref{fig:overview-arguzz}:
(1)~circuit generation, (2)~circuit transformation,
(3)~circuit-to-Rust translation, (4)~product-program generation,
(5)~input generation, (6)~normal VM execution and trace collection,
and (7)~VM execution with malicious prover. We describe each step at a
high level below and provide technical details in
\secref{sec:approach}.

\paragraph{Step 1: Circuit generation.}
We begin by generating a random circuit in \il. This circuit
represents a computation including typical control-flow patterns and
arithmetic operations used in zero-knowledge applications. Unlike
prior work~\cite{HochrainerIsychev2025}, \tool extends this step (and
\il) to optionally include inline-assembly constructs that target
specific instructions supported by target zkVMs. This ensures that
even low-level or uncommon operations are exercised during testing.
\figref{fig:ex-il1} shows an example circuit, denoted $C_1$, generated
in this step. Lines~\ref{l:cins}--\ref{l:couts} declare the circuit
inputs and outputs, and line~\ref{l:cout} computes the value of the
output using a basic arithmetic expression.

\paragraph{Step 2: Circuit transformation.}
We use a similar set of semantics-preserving transformations as
\circuzz~\cite{HochrainerIsychev2025}. These include transformations
based on algebraic identities, such as commutativity, associativity,
distributivity, and De Morgan's laws. They apply to logical, bitwise,
and arithmetic operations alike.
We disable transformations that are specific to field arithmetic and
not applicable to general-purpose zkVMs. In addition, we enrich the
transformation set with new rules to broaden coverage. A complete list
of included transformations is provided in
\appref{sect:appendix:rules}.

In practice, we stack multiple transformations on the original circuit
to produce a transformed circuit that is syntactically different but
semantically equivalent.
\figref{fig:ex-il2} shows the result of applying two transformations
to the original circuit $C_1$. First, we apply the commutativity of
addition (rule \code{comm-add} in \appref{sect:appendix:rules}) to
obtain output expression \code{(a \% (c + b))}, followed by addition
with the identity element (rule \code{zero-add-con} in
\appref{sect:appendix:rules}), which adds zero to \code{c}. Obviously,
the resulting circuit, $C_2$, remains semantically equivalent to
$C_1$.

In general, metamorphic transformations serve as an \emph{oracle} for
the expected behavior of the zkVM: the original and transformed
circuits should produce the same output when executed---any divergence
indicates a bug.

\paragraph{Step 3: Circuit-to-Rust translation.}
The original and transformed \il circuits are then independently
compiled to Rust functions. This translation preserves the semantics
of each circuit and includes any inline assembly specified during
generation.
\figsref{fig:ex-rust1} and \ref{fig:ex-rust2} show the Rust
translations of $C_1$ and $C_2$, respectively. Each function takes the
circuit inputs, computes the output using standard Rust syntax, and
returns the result.

\paragraph{Step 4: Product-program generation.}
Next, we merge the Rust functions into a single \emph{product
program}. This program executes the functions and checks that their
outputs match. The structure is inspired by work on
hyperproperty~\cite{ClarksonSchneider2008}
reasoning~\cite{BartheDArgenio2004,TerauchiAiken2005,BartheCrespo2011},
where product constructions are used to reason about relationships
between multiple program executions. In our setting, the product
program computes a known, expected output only if both function
executions behave identically. This enables \tool to detect behavioral
mismatches while avoiding the overhead of executing each function as a
separate program.

\figref{fig:prod-prog} shows the product program (in Rust) generated
by \tool using the functions of \figsref{fig:ex-rust1} and
\ref{fig:ex-rust2}. On lines~\ref{l:call-c1}--\ref{l:call-c2}, it
calls the functions, and on line~\ref{l:outs}, it compares their
outputs. If they differ, the program returns a special value
\code{OOPS} (line~\ref{l:outs-no}); otherwise, it returns a
\code{SUCCESS} value (line~\ref{l:outs-yes}).

\begin{figure}[t!]
\begin{lstlisting}[style=basic]
const OOPS: u32 = 0x0; @\label{l:oops}@
const SUCCESS: u32 = 0xC0FFEE; @\label{l:success}@

// circuit c1 as Rust function
fn c1(a: u32, b: u32, c: u32) -> u32 {
  a % (b + c)
}

// circuit c2 as Rust function
fn c2(a: u32, b: u32, c: u32) -> u32 {
  (a % ((c + 0) + b))
}

// VM entry point
[zkvm::entry(main)]
fn main(a: u32, b: u32, c: u32) -> u32 {
  let c1_out = c1(a, b, c); @\label{l:call-c1}@
  let c2_out = c2(a, b, c); @\label{l:call-c2}@

  // check if violation occurred
  if c1_out != c2_out { @\label{l:outs}@
    OOPS    // unexpected result @\label{l:outs-no}@
  } else {
    SUCCESS // expected result @\label{l:outs-yes}@
  }
}
\end{lstlisting}
\caption{Product program in Rust generated by \tool using the
  functions of \figsref{fig:ex-rust1} and \ref{fig:ex-rust2}.}
\label{fig:prod-prog}
\end{figure}

\paragraph{Step 5: Input generation.}
We generate random private and public inputs for the product
program. These inputs are used for subsequently executing the product
program in the VM.
For example, for the program of \figref{fig:prod-prog}, \tool might
randomly generate the values \code{7} for \code{a}, \code{3} for
\code{b}, and \code{2} for \code{c}.

\paragraph{Step 6: Normal VM execution and trace collection.}
Given the generated inputs, the product program is executed inside the
unmodified zkVM. If it crashes or produces an output different from
the expected one, we flag this as a potential soundness or
completeness bug.
During this execution, we also collect the trace of the product
program, which records the sequence of executed instructions. This
trace is used in the next step to guide fault injection: by
identifying which instructions the VM executed, we can target faults
at the corresponding points in the VM's instruction-handling logic.

For example, assume that, for input values \code{7}, \code{3}, and
\code{2}, the product program in \figref{fig:prod-prog} crashes; this
indicates a completeness bug. Now, assume that it returns \code{OOPS};
this may indicate a soundness bug, a completeness bug, or an issue
that manifests as both. Finally, assume that it returns
\code{SUCCESS}; no bug is detected.
If no bug is detected, we observe that the \riscv
\code{remu} instruction (unsigned remainder) is executed twice (as
part of computing the return value of each Rust function). Knowing
this, allows us to target the VM's implementation of \code{remu} in
a fault-injected run, increasing the chances of exposing soundness
bugs related to that specific operation.

\paragraph{Step 7: VM execution with malicious prover.}
Finally, we re-run the product program with the same inputs and mimic
prover misbehavior by injecting faults directly into the zkVM. This
process is guided by the trace collected in the previous step:
we target the VM's handling of instructions that were actually
executed. For example, we may modify their operands or output
values. This allows us to explore how the verifier behaves in the
presence of a malicious prover. If the fault injection causes the
product program to return an incorrect output (i.e., \code{OOPS}) but
the proof still verifies successfully, \tool reports a soundness
bug. The verifier accepting an invalid trace indicates that the
constraints are underspecified.

\begin{figure}[t!]
\begin{lstlisting}[style=basic]
  if c1_out != c2_out {
    OOPS // unexpected result
  else if c2_out != c3_out {
    OOPS // unexpected result
  } else {
    SUCCESS // expected result
  }
}
\end{lstlisting}
\caption{Part of product program in Rust generated from three
  semantically equivalent circuits.}
\label{fig:prod-prog3}
\end{figure}

Recall that, in our example, the trace collected in the previous step
shows that instruction \code{remu rd, rs1, rs2} is executed twice,
where \code{rd} is the destination register, \code{rs1} the dividend,
and \code{rs2} the divisor. For our input values, \code{rs1 = 7},
\code{rs2 = 5}, \code{rd} is assigned the value \code{7 \% 5}. In this
step, we re-run the product program with the same inputs, and to mimic
a malicious prover, we automatically choose to inject a fault that
modifies the behavior of one of the executed \code{remu}
instructions. Specifically, we replace the divisor \code{rs2} with
\code{rs1}, causing the VM to compute \code{remu rd, rs1, rs1}
instead. This yields \code{7 \% 7 = 0} as the return value of one of
the Rust functions, and in turn, the product program returns
\code{OOPS}. However, due to a missing constraint in \risc, the
generated proof still verifies successfully. After minimizing the
product program, we are able to verify that \code{7 \% 5 = 0}!

We reported this soundness bug to the \risc developers, who classified
the issue as critical and awarded a \$50,000 bounty. Importantly, the
bug was not limited to the \code{remu} instruction; it affected any
instruction that used three register operands, such as \code{divu},
due to missing checks in the constraint system. \tool exposed multiple
such cases. The issue was subsequently patched with changes to eleven
files in
\zirgen\footnote{\url{https://github.com/risc0/zirgen/pull/238}},
\risc's constraint-system implementation, and 32 files in the
\risc\footnote{\url{https://github.com/risc0/risc0/pull/3181}} zkVM
implementation. A new release was issued, and all clients were
migrated to the updated version.

\bigskip
\medskip
Note that \tool performs two loops, as shown in \figref{fig:overview-arguzz}. The
first one (steps 2--4) generates new transformed circuits to exercise
the constraint system in different ways. The corresponding Rust
functions are all merged into the same product program for
comparison. For example, we could generate a third transformed circuit
and corresponding Rust function \code{c3} and merge it in the product
program as shown \figref{fig:prod-prog3}. This allows us to compare
multiple semantically equivalent functions in a single execution and
detect inconsistencies across any of them.
The other loop (steps 5--7) tests each product program across
different inputs and fault injections. In general, these loops
increase the likelihood of finding bugs with each product program.

%%% Local Variables:
%%% mode: latex
%%% TeX-master: "main"
%%% End:

%!TEX root = main.tex

%-------------------------------------------------------------------------------
\section{Approach}
\label{sec:approach}
%-------------------------------------------------------------------------------

As outlined in the previous section, our method combines two
complementary ideas: a new variant of metamorphic testing and fault
injection. Our metamorphic-testing variant provides an effective
strategy for generating product programs with a known output; this
allows detecting soundness and completeness bugs. On the other hand,
fault injection simulates misbehavior by the prover to detect
soundness bugs due to overly weak constraints.
Note, however, that while our design leverages the synergy of these
two ideas, they are technically orthogonal. In particular, any
technique that produces programs with known outputs could be used in
place of our product-program generator.

\paragraph{\tool workflow.}
As discussed, the \tool workflow consists of seven steps. In this
section, we focus on the three more technically involved components of
the approach--(1)~circuit generation, (4)~product-program
generation, and (7)~VM execution with malicious prover. The remaining
steps are conceptually straightforward and summarized below:
\begin{itemize}
\item \textbf{Step 2: Circuit transformation.} This step applies
  semantics-preserving rewrites to the original circuit to produce a
  transformed variant. We reuse most transformations from \circuzz,
  omitting those tailored to field arithmetic (irrelevant for zkVMs),
  and enriching the set with new rules (see \appref{sect:appendix:rules}).

\item \textbf{Step 3: Circuit-to-Rust translation.} Each \il circuit
  (original and transformed) is compiled into a standalone Rust
  function. This translation is direct and preserves circuit
  semantics, including inline assembly.

\item \textbf{Step 5: Input generation.} Inputs are generated using a
  blackbox-fuzzing strategy guided by the type signatures of the Rust
  functions. We maintain a configurable set of constants that include
  interesting boundary values (e.g., $0$, $1$, $-1$, maximum or
  minimum integers, etc.) to increase the likelihood of triggering
  edge cases.

\item \textbf{Step 6: Normal VM execution and trace collection.} The
  product program is executed inside the unmodified zkVM. We collect
  the trace record to determine which parts of the VM are
  exercised. This information is used to guide fault injection in the
  next step.
\end{itemize}

%-------------------------------------------------------------------------------
\subsection{Circuit Generation}
\label{sec:step1}
%-------------------------------------------------------------------------------

\tool begins by generating a circuit expressed in \il, the
intermediate language introduced in
\circuzz~\cite{HochrainerIsychev2025} for testing ZK pipelines. This
circuit forms the basis for metamorphic transformations and subsequent
execution in the zkVM.

To systematically explore the VM's behavior, we extend \il with custom
functions that emit inline \riscv assembly. This allows \tool to
explicitly include specific instructions---such as \code{mulhsu},
which computes the upper half of the product of two integers---in the
generated circuit. Such an extension is critical for ensuring broad
coverage across the instruction set supported by each zkVM.
As an example, the \il circuit in \figref{fig:ex-inline-il} calls the
custom \code{mulhsu} function. Its Rust translation, shown in
\figref{fig:ex-inline-rust}, implements this function using a macro
and inline assembly on lines~\ref{line:asm-begin}--\ref{line:asm-end}.

\begin{figure}[t!]
\begin{subfigure}[b]{0.5\textwidth}
\begin{lstlisting}[style=il]
inputs : a, b, c
outputs: out
out = mulhsu(a, (b + c)) @\label{l:mulhsu}@
\end{lstlisting}
\vspace{-0.5em}
\caption{Circuit $C$ in \il.}
\vspace{1em}
\label{fig:ex-inline-il}
\end{subfigure}\\
\begin{subfigure}[b]{.5\textwidth}
\begin{lstlisting}[style=basic,morekeywords={out}]
fn c(a: u32, b: u32, c: u32) -> u32 {
  macro_rules! mulhsu { @\label{line:asm-begin}@
    (@\$@a:expr, @\$@b:expr) => {{
      let result: u32;
      unsafe {
        core::arch::asm!(
          "mulhsu {result}, {a}, {b}",
          result = out(reg) result,
          a = in(reg) @\$@a,
          b = in(reg) @\$@b,
        );
      }
      result
    }}
  } @\label{line:asm-end}@
  mulhsu!(a, (b + c))
}
\end{lstlisting}
\vspace{-1em}
\caption{Circuit $C$ as a Rust function.}
\label{fig:ex-inline-rust}
\end{subfigure}
\caption{Example \il circuit and Rust function generated by \tool
  using the inline-assembly extension.}
\label{fig:inline-assembly}
\end{figure}

%-------------------------------------------------------------------------------
\subsection{Product-Program Generation}
\label{sec:step4}
%-------------------------------------------------------------------------------

Our fault-injection mechanism requires programs with known outputs to
reliably detect soundness bugs. However, generating programs of
configurable complexity with predictable outcomes is challenging,
especially when targeting a diverse set of low-level instructions
supported by zkVMs.

\paragraph{Metamorphic testing.}
Metamorphic testing~\cite{ChenCheung1998} provides an effective
solution to this problem. It is a well established technique used for
testing complex software systems, including
compilers~\cite{ChenPatra2020}, program analyzers~(e.g.,
\cite{ZhangSu2019,MansurChristakis2020,WintererZhang2020-Fusion,MansurChristakis2021,MansurWuestholz2023,ZhangPei2023,MordahlZhang2023,ZhangPei2024,HeDi2024,KaindlstorferIsychev2024}),
and ZK pipelines~\cite{HochrainerIsychev2025}. In such contexts,
metamorphic testing generates two semantically equivalent yet
syntactically different programs whose outputs must
match. Specifically, the \circuzz fuzzer for ZK pipelines generates an
original circuit in \il and derives a transformed variant by applying
a random sequence of semantics-preserving rewrites.

While metamorphic testing guarantees output equivalence between the
original and transformed programs, their actual outputs are not known
a priori. Therefore, traditional metamorphic testing executes both
programs separately and compares their outputs externally.

\paragraph{Metamorphic oracles as product programs.}
In \tool, we generate a single product program from the original and
transformed Rust functions generated in the previous step. Instead of
running separate executions and performing external checks, our
product program internally computes the outputs of all functions and
directly compares them. Specifically, it produces one of two outcomes
(see \figref{fig:prod-prog} for an example): \code{SUCCESS},
indicating all outputs match, or \code{OOPS}, indicating a
mismatch. Thus, we effectively encode the metamorphic oracle directly
into the product program. This often eliminates execution
overhead and elegantly solves the requirement of knowing the output
beforehand---the expected output is always \code{SUCCESS} unless the
zkVM under test is buggy.

Regarding execution overhead, product programs often improve test
throughput, mostly by reducing prover costs. Proving dominates runtime
and typically scales super-linearly with trace size. Moreover, many
zkVMs pad traces to the nearest power of two, which amplifies
costs. If two executions are run separately, each incurs its own
padding overhead. With a product program, padding is applied only
once. For instance, a single trace of size 500 (padded to 512) is
cheaper to prove than running two separate traces of sizes 280 and
180, which would be padded to 512 and 256 respectively.
This effect becomes even more pronounced when bundling more than two
Rust functions into a single product program, since multiple padding
overheads are avoided.

The concept of a product program was originally introduced in the
context of hyperproperty~\cite{ClarksonSchneider2008}
reasoning~\cite{BartheDArgenio2004,TerauchiAiken2005,BartheCrespo2011}. Rather
than invoking multiple program variants externally to check a
hyperproperty, product programs encode the multiple executions
internally and perform the necessary comparisons within a single
combined program. This internalization enables hyperproperty checking
without requiring repeated calls to an external oracle. For example,
such reductions allow standard program-verification tools---designed
for single-execution properties---to be applied to relational or
multi-trace properties.

Metamorphic testing can be viewed as a form of hyperproperty
reasoning, typically targeting $2$-safety properties--- for such
properties, a failing test consists of two executions. Note that, in \tool,
metamorphic testing checks $k$-safety properties because of the first
loop in \figref{fig:overview-arguzz} (steps 2--4), which produces and compares
$k$ transformed circuits within the same product program. While
product programs have a rich history in program verification, we are
not aware of prior work that applies them directly to metamorphic
testing.

%-------------------------------------------------------------------------------
\subsection{VM Execution with Malicious Prover}
\label{sec:step7}
%-------------------------------------------------------------------------------

Fault injection is a well established technique for uncovering bugs in
software systems. The core idea is that injected faults should trigger
an appropriate system response---ideally failing fast and gracefully,
rather than causing silent or progressive system corruption. For
example, chaos engineering, popularized by Netflix, tests the
resilience of distributed systems by introducing random faults such as
network outages.

\paragraph{The zkVM threat model.}
The threat model for zkVMs mirrors that of other zero-knowledge
systems: the verifier is the only trusted component, while the prover
is potentially adversarial. In other words, an attacker may use a
malicious execution environment to generate an invalid trace and a
corresponding proof, and it is the verifier's responsibility to detect
and reject such proofs.

\paragraph{Designing effective fault injection.}
To test this trust boundary, we present the first fault-injection
mechanism for zkVMs. In particular, we simulate a malicious prover by
injecting faults into the VM's execution logic. If the product program
returns an unexpected output but the verifier is successfully deceived
into accepting the resulting proof, this indicates a soundness bug.

Our fault-injection mechanism deliberately encourages a ``ripple
effect'' by introducing faults that propagate naturally through the
execution trace. We inject a fault into the execution of a single
instruction, e.g., by adding $1$ to the result of a
multiplication. The modified value is written to a register, and any
subsequent instructions that read from that register propagate
the altered result. This allows the fault to cascade through the trace
in a way that respects normal data dependencies, increasing the
likelihood that the resulting proof verifies.

With this design, the execution trace remains valid up to the point of
injection. Immediately after the fault, a small, localized
inconsistency may be introduced, but the remainder of the trace
becomes consistent again---this time with respect to the faulty
state. Such a localized inconsistency is more likely to evade
detection if the corresponding constraints are insufficiently
precise for those specific points in the trace.

An alternative design we initially considered was to generate a valid
trace record and then randomly flip bits to test whether the verifier
would still accept the corresponding proof. However, bit flips are not
guaranteed to produce invalid traces. For example, consider a trace
computing \code{42 * 0}: flipping the operand from \code{42} to
\code{43} would still yield the same valid result of zero. By
contrast, our fault-injection approach leverages both the verifier's
output and the program's expected output to determine whether the
proofs that verify are about invalid traces and avoid such false
positives.

\begin{figure}[t!]
\begin{lstlisting}[style=basic]
fn execute(/* ... */) -> /* ... */ {
  let mut i: Instruction;

  // < ZKVM INSTRUCTION DECODING > @\label{line:decode}@

  // instruction-modification injection @\label{line:inject-start}@
  if is_injection_enabled() && @\label{line:inject-flag}@
     is_injection_type("INSTR_MOD") && @\label{line:inject-type}@
     is_injection_step() @\label{line:inject-step}@
  {
    let new_i = fuzzer::new_instr(&i); @\label{line:fuzzer-instr}@
    i = new_i;
  } @\label{line:inject-end}@

  // < ZKVM INSTRUCTION EXECUTION > @\label{line:exec}@

  // increments the injection step counter
  fuzzer::step(); @\label{line:step-incr}@

  // ...
}
\end{lstlisting}
\caption{Generic instruction-modification injection performed by
  \tool.}
\label{fig:inject}
\end{figure}

\paragraph{Fault-injection types.}
The fault-injection component is implemented by augmenting the zkVM's
execution stage with custom logic. We define several injection
\emph{types} that describe how a fault is applied. For example, an
instruction-modification injection may change the operation, the
output value, or a register operand---such as altering the divisor
register in a \code{remu} instruction (see
\secref{sec:overview})---while a memory-modification injection
writes arbitrary values into memory.

Since some injection types are tailored to the internal design of a
specific VM, we focus our discussion on a generic type that can be
applied universally across a wide range of zkVMs, namely, the
instruction-modification injection. \figref{fig:inject} shows how we
modify the execution stage of a zkVM to implement this injection
type. At a high level, an instruction is decoded
(line~\ref{line:decode}), the instruction is replaced by a new, fuzzed
variant if certain conditions are met
(lines~\ref{line:inject-start}--\ref{line:inject-end}), and the
instruction is executed (line~\ref{line:exec}).

To decide where to inject, we currently use a custom, global step
counter that increments with each executed instruction and triggers
the injection when it matches a target step. The target step is
selected by a fault-injection scheduler, described later in this
section. In the figure, the step counter is checked on
line~\ref{line:inject-step}. This mechanism is not fundamental: it can
be replaced with other targeting schemes, such as using the cycle
count or program counter.  After the instruction execution, the
counter is incremented on line~\ref{line:step-incr}.

To avoid accidental injections during normal execution, all injected
code is guarded by a global injection flag, which is set only for
fault-injection runs (line~\ref{line:inject-flag}). A second check
verifies that the injection type matches the target one
(line~\ref{line:inject-type})---the target injection type is chosen
randomly by the fuzzer. Finally, a third check ensures that the
current step matches the target injection point
(line~\ref{line:inject-step}) as discussed earlier.

Note that certain injection types rely on randomized values---for
example, generating a random operation, value, or operand---to
increase behavioral diversity across test runs. These values are
provided by the fuzzer, e.g., line~\ref{line:fuzzer-instr}
randomly fuzzes the current instruction to generate a new variant.

While the details of fault-injection code vary across zkVMs due to
internal architectural differences, the instruction-modification
injection type (shown in \figref{fig:inject}) is implemented in all
the VMs we tested. Other injection types include modifying the program
counter or altering the output of an operation before it is written
to a register or memory. In some cases, we developed custom injection
types tailored to a specific VM. For instance, \openvm adopts a
chip-based design, where each chip is a modular execution unit
responsible for individual operations or families of operations; in
this setting, we created injection types targeted at particular chips.
Notably, all the soundness bugs uncovered by \tool were triggered by
the instruction-modification injection. A complete list of the
injection types in \tool can be found in
\appref{sect:appendix:injections}.

Ensuring that injected faults actually take effect is crucial. Some
faults may otherwise be blocked by safety checks in the prover's
code. To prevent this, we replace built-in assertion and
panic macros in the VM with custom versions. When the injection flag is set,
these macros disable selected runtime checks, allowing injections to
proceed without being prematurely aborted and to propagate through
execution.

\paragraph{Fault-injection scheduler.}
A naive version of our fault-injection strategy could inject faults at
random points in the execution trace. However, in real programs,
certain instructions---such as memory reads or additions---occur far
more frequently than others. As a result, purely random injection
would disproportionately target common instructions, leaving rarer
instructions undertested.

To address this imbalance, we implement a fairer fault-injection
scheduler that aims to uniformly cover all available \riscv
instructions. The fuzzer maintains a count of how often each
instruction has been targeted for injection. When analyzing the trace
collected during normal execution, we identify the least frequently
injected instructions and randomly select one of them. The scheduler
then injects a fault at that instruction. If the selected instruction
appears multiple times in the trace, the injection point is chosen
uniformly at random among its occurrences.

\paragraph{Soundness-bug detection.}
To avoid generating false positives, \tool does not report a bug
solely because the verifier accepts a proof after fault injection. We
additionally require that the product-program output
changes from \code{SUCCESS} to \code{OOPS}.
If the program crashes or produces the expected output, we cannot draw
any definitive conclusions. The crash may have been caused by an
unrelated side effect of the injection, or the fault may have failed
to influence the execution.

For instance, we observed that some injections---such as adding $2$ to
the first operand of a modulo operation with $2$---do not affect the
output or trace in a detectable way. By requiring a change in the
output of the product program, we ensure that the injected fault has a
concrete, observable impact on the execution trace---one that the
verifier should reject.

%%% Local Variables:
%%% mode: latex
%%% TeX-master: "main"
%%% End:

%!TEX root = main.tex

%-------------------------------------------------------------------------------
\section{Experimental Evaluation}
\label{sec:evaluation}
%-------------------------------------------------------------------------------

We evaluate \tool by testing six popular zkVMs. In our
evaluation, we address the following research questions:
\begin{description}
\item[RQ1:] How effective is \tool in detecting soundness and
  completeness bugs in zkVMs?
\item[RQ2:] What are characteristics of the detected bugs?
\item[RQ3:] How efficient is \tool?
\item[RQ4:] How effective is the inline-assembly extension?
\item[RQ5:] How effective is the fault-injection scheduler?
\item[RQ6:] What is the impact of the instruction-modification
  fault-injection type?
\end{description}

%-------------------------------------------------------------------------------
\subsection{Experimental Setup}
\label{subsec:setup}
%-------------------------------------------------------------------------------

\paragraph{zkVM selection.}
To ensure that our evaluation focuses on zkVMs with real-world impact,
we consulted with the Ethereum Foundation, who provided us with a list
of mature zkVM implementations. From this list, we selected six zkVMs
that are actively maintained and representative of the state of the
art. This selection balances diversity of design choices with
practical relevance.

\paragraph{Experiments in the wild.}
The primary goal of these experiments was to assess \tool's
effectiveness in discovering previously unknown bugs in mature
zkVMs. We began testing \risc and \nexus in March 2025 and gradually
extended \tool to support additional zkVMs. Support for \spOne was
added in May 2025, followed by \jolt, \openvm, and \pico in June
2025. Once a bug was discovered in a zkVM, we typically paused fuzzing
that system until the issue was fixed to avoid reporting
duplicates. For all zkVMs, we tested their respective main branches.

\paragraph{Controlled experiments.}
In our controlled experiments, we evaluated \tool using two
configurations.

The first focused on \emph{bug refinding}, i.e., assessing whether the
reported bugs can be rediscovered. In this setup, we launched $5$
fuzzing campaigns, each with a different numeric seed to initialize
the fuzzer's random-number generator. Each campaign ran on $4$ CPUs
and was limited to 24h. To obtain transformed circuits, we applied
between $1$ and $4$ stacked metamorphic transformations. Each product
program bundled between $2$ and $10$ Rust functions, corresponding to
the loop in steps~2--4 of \figref{fig:overview-arguzz}. We executed
each product program under $3$ different combinations of inputs and
fault injections, corresponding to the loop in steps~5--7 of
\figref{fig:overview-arguzz}.
For tractability of the experiments, we enabled only the
instruction-modification injection, which is responsible for all
detected soundness bugs.
To confirm that a bug found with this configuration was the same as the
one originally reported, we applied the fix provided by the developers
and checked that the bug disappeared.

The second configuration focused on \emph{feature evaluation}, i.e.,
quantifying the benefits of \tool's design contributions such as the
inline-assembly extension, the fault-injection scheduler, and other
components. This setup used the same parameters as the bug-refinding
configuration, with the exception that each campaign was run with one
random seed.

\paragraph{Hardware.}
We performed all experiments on a machine with an AMD EPYC 9474F CPU @
3.60GHz and 1.5TB of memory, running Debian GNU/Linux 12 (bookworm).

%-------------------------------------------------------------------------------
\subsection{Experimental Results}
\label{subsec:results}
%-------------------------------------------------------------------------------

\begin{table*}[t]
\caption{Unique bugs detected by \tool.}
\centering
\begin{tabular}{c|c|c|c|c|l}
\begin{tabular}[c]{@{}c@{}}\textbf{Bug ID}\end{tabular} & \textbf{zkVM}
& \textbf{Fix ID} & \textbf{Type} & \textbf{Oracle} & \textbf{Description}\\
\hline
\bug{1}  & \risc  & \fix{A},\fix{B} & soundness    & FI & Missing constraint in three-register instructions\\
\bug{2}  & \risc  & \fix{C}         & completeness & MT & Off-by-one error in cycle-counting logic\\
\hline
\bug{3}  & \nexus & \fix{D}         & soundness    & FI & Unconstrained store operand in load-store instructions\\
\bug{4}  & \nexus & \fix{E}         & completeness & MT & Out-of-bounds panic due to memory size misestimation\\
\bug{5}  & \nexus & \fix{F}         & completeness & MT & Carry overflow in multiplication extension\\
\hline
\bug{6}  & \jolt  & \fix{G}         & soundness    & FI & Unconstrained immediate operand in \code{lui}\\
\bug{7}  & \jolt  & \fix{H}         & completeness & MT & Incorrect RAM size calculation\\
\bug{8}  & \jolt  & \fix{I}         & completeness & MT & Out-of-bounds panic for high-address bytecode\\
\bug{9}  & \jolt  & \fix{J}         & completeness & MT & Dory-commitment failure for traces shorter than 256 cycles\\
\bug{10} & \jolt  & \fix{K}         & completeness & MT & Sumcheck-verification failure for \code{mulhsu}\\
\bug{11} & \jolt  & -               & completeness & MT & Sumcheck-verification failure for inline \code{div} and \code{rem}\\
\end{tabular}
\label{tab:bugs}
\end{table*}

\paragraph{RQ1: Effectiveness of \tool.}
\tabref{tab:bugs} summarizes all previously unknown, unique bugs
uncovered by \tool in our in-the-wild experiments. The first column
assigns each bug an identifier (ID) and, where available, links to the
corresponding public bug report. The second column lists the zkVM in which the bug was
discovered---importantly, we found issues in three separate zkVMs. The
third column links to the pull requests containing the corresponding
patches. Note that bug\bug{11} remains unfixed, as it was discovered
only during the week of the paper-submission deadline. The developers
initially pushed a
fix\footnote{\url{https://github.com/a16z/jolt/pull/898}}, but later
reverted it after realizing that it did not fully resolve the
issue. The fourth column classifies each bug by its impact,
distinguishing whether it affected soundness, completeness, or
both. The fifth column specifies the oracle that exposed the bug:
``MT'' (metamorphic testing) indicates that the product program either
crashed when executed on the unmodified VM or produced an unexpected
output; ``FI'' (fault injection) means that the product program was
executed on a malicious prover, produced an unexpected output, and yet
the verifier was deceived into accepting the resulting proof. The last
column provides a brief description of each bug.

\emph{In total, \tool uncovered eleven unique, previously
unknown bugs across three distinct zkVMs.} This result is significant
given that these are mature zkVMs, routinely subjected to audits and
developed under rigorous engineering practices. Notably, both
bugs identified in \risc were acknowledged with bounties: bug\bug{1}
was awarded \$50,000 and bug\bug{2} \$1,000. In contrast, \nexus and
\jolt did not operate bounty programs, though they promptly addressed
the reported vulnerabilities. \emph{Of the found bugs, the
eight completeness bugs were detected with metamorphic testing, while the three soundness bugs were revealed
through fault injection.}
This highlights the complementary strengths
of the two techniques: metamorphic testing was more prolific in
uncovering completeness bugs, whereas fault injection revealed the
most critical soundness vulnerabilities.

\paragraph{RQ2: Detected bugs.}
To better understand the types of issues exposed by \tool, we now
examine several of the discovered bugs in more detail.

Bug\bug{1}, which we already introduced in \secref{sec:overview}, was
present in \risc versions 2.0.0, 2.0.1, and 2.0.2. It affected all
instructions with three register operands, such as \code{divu} and
\code{remu}. The root cause was a missing constraint that caused the
VM to fail to distinguish between the values of the first and second
operand registers. As a result, proofs generated by a malicious prover
could go undetected by the verifier, leading to a critical soundness
bug. Addressing this bug required coordinated fixes across two
repositories: not only the main \risc zkVM implementation, but also
the \zirgen repository, which contains \risc's constraint-system
implementation.

Bug\bug{2} was due to a subtle off-by-one error in a prover component
responsible for counting execution cycles: the final processing step
was not included in the total. At first, this mistake was masked by an
unrelated boundary-condition issue, preventing immediate
failures. However, the incorrect count eventually propagated to later
stages, where it caused downstream validation to fail.
\tool uncovered this bug by fuzzing programs of varying sizes, which
triggered the specific conditions needed for the error to
manifest. The developers noted they were impressed that fuzzing
revealed it, and we received a \$1,000 bounty for the report.

Bug\bug{3} was a missing constraint in \nexus's handling of
load-store instructions, which allowed a malicious prover to exploit an
unconstrained memory write. Specifically, the lower bits of the
register holding the store value were not properly constrained for the
\riscv store instructions \code{sw}, \code{sh}, and \code{sb}. As a
result, a prover could alter these bits without detection.
\tool uncovered this issue with the instruction-modification injection
(see \figref{fig:inject}). In this case, the injection changed the
second operand---the store value---to reference an arbitrary
register. The flaw surfaced once the value was read back from memory
and enabled unsoundly verifying $2^3 \oplus 2^3 = 1$.

Bug\bug{4} was caused by an incorrect estimation of the size of
touched or initialized memory, which led to an out-of-bounds panic
inside the prover for certain programs. After our report, the
developers promptly issued a dedicated patch in \nexus version 0.3.1
to resolve the issue. Once again, the diverse range of programs
generated by \tool revealed an untested edge case, exposing a
completeness bug.

Bug\bug{5} occurred after the introduction of \riscv's multiplication
extension in \nexus. It manifested as a panic indicating that
carry-flag bounds had been exceeded, and it affected multiplication,
division, and remainder operations. In this case, \tool was essential
not only for generating a suitable program but also for
producing the input arguments required to trigger the
bug. Fixing the issue required a rework of the carry logic across all
\riscv multiplication operations in \nexus.

Bug\bug{6} was caused by a missing constraint in the \riscv \code{lui}
instruction in \jolt (v0.1.0). The \code{lui} instruction is supposed to load a
16-bit immediate value into the upper 16 bits of a target register.
Due to a missing constraint, a malicious prover could arbitrarily
manipulate the immediate operand and thereby control the instruction's
output. Similar to bug\bug{3}, \tool discovered this issue through an
instruction-modification injection (see \figref{fig:inject}), which in
this case manipulated the immediate operand.
Notably, all three soundness bugs discovered by \tool were detected
using the instruction-modification injection type.

In addition to the previously discussed issues, \tool uncovered
five further completeness bugs in \jolt (v0.2.0). Bug\bug{7} stemmed from an
incorrect RAM size calculation that prevented certain program bytecode
from fitting in memory. Bug\bug{8} triggered an out-of-bounds panic
whenever program bytecode was placed at higher memory addresses than
those accessed during execution. Bug\bug{9} was caused by the new Dory
commitment~\cite{Lee2020} implementation, which failed on short traces
containing fewer than 256 cycles. Bug\bug{10} revealed a
sumcheck-verification failure for the \code{mulhsu} instruction; note
that \il circuits contained this instruction thanks to our
inline-assembly extension (see \secref{sec:step1}). Finally,
bug\bug{11} caused a sumcheck failure when verifying inline \code{div}
and \code{rem} instructions.

Beyond these zkVM bugs, \tool also uncovered a Rust compiler bug as a
by-product while testing \risc. The issue was present in Rust version
1.80, which \risc used to compile input programs. A miscompilation
caused a boolean expression to evaluate incorrectly, leading to both
soundness and completeness issues. Further details about the affected
expression can be found in the corresponding \risc bug
report\footnote{\url{https://github.com/risc0/risc0/issues/2878}}.

\paragraph{RQ3: Efficiency of \tool.}
In this research question, we evaluate the efficiency of \tool along
three dimensions.

We first conducted a bug-refinding experiment to measure how quickly
\tool can rediscover previously reported bugs. Bug\bug{11} was not
included as it is not yet fixed. Out of the ten remaining bugs, \tool
successfully rediscovered six (bugs\bug{1}, \bug{3}, \bug{5}, \bug{6},
\bug{9}, and \bug{10}), including all soundness bugs, within a median
time of 13h. In fact, two of these were refound in under 20min and
five within 6h. The remaining completeness bugs require longer (days)
as they depend on very specific conditions to manifest (see
RQ2). Detailed per-bug results are reported in
\appref{sect:appendix:results}.
These results are encouraging: all soundness bugs, which are the most
critical from a security perspective, would have been discovered with
only a few hours of fuzzing despite the size, complexity, and slow
execution speeds of zkVMs. Completeness bugs\bug{2}, \bug{4}, \bug{7},
and \bug{8} are inherently harder to hit, as they only arise under
narrowly defined circumstances, which naturally prolongs rediscovery.

Second, we measure the runtime overhead of enabling fault injection in
\tool, shown in \tabref{tab:runtime}. For this experiment, we use the
feature-evaluation configuration and run \tool both with and without
fault injection enabled. The first column lists the zkVM, the second
shows the number of tested product programs with injection enabled,
and the third shows the number of tested programs with injection disabled (i.e.,
omitting step~7 in \figref{fig:overview-arguzz}). As the table shows,
throughput increases substantially when disabling fault injection,
ranging from 25.6\% for \openvm to 98.4\% for \pico. The increase does
not reach a full 2x, however, despite omitting the additional VM run
in step~7. This is expected for two reasons: (1)~the earlier \tool
steps introduce common overhead, and (2)~injected runs often terminate
early (see RQ6), reducing their runtime.

\begin{table}
\centering
\begin{tabular}{c|c|c}
  \textbf{zkVM} &
  \textbf{\makecell{Programs\\(injection on)}} &
  \textbf{\makecell{Programs\\(injection off)}} \\
  \hline
  \risc   & 1468 & 2294 \\ % +56.3%
  \nexus  &  788 & 1376 \\ % +74.6%  
  \jolt   &  377 &  503 \\ % +33.4%
  \spOne  &  494 &  880 \\ % +78.1%  
  \openvm & 1267 & 1591 \\ % +25.6%
  \pico   &  549 & 1089 \\ % +98.4%
\end{tabular}
\caption{Number of product programs executed in each zkVM with and
  without fault injection enabled.}
\label{tab:runtime}
\end{table}

Third, we evaluate the impact of product-program size on VM execution
time, since running the VM is the slowest component in our
workflow. For this, we use the feature-evaluation configuration and
disable fault injection to avoid confounding effects on runtime. We
observed that execution times typically fall within a very narrow
range, even though program sizes vary widely. In particular, we did
not observe a linear relationship between program size and execution
time. For this reason, the number of semantically equivalent functions
bundled into a product program has little effect on runtime. This is
also why our metamorphic-testing variant with product programs tends
to be more efficient than traditional metamorphic
testing. \appref{sect:appendix:results} provides plots illustrating
this trend.

\paragraph{RQ4: Effectiveness of inline-assembly extension.}
We now evaluate the effectiveness of our inline-assembly extension
(see \secref{sec:step1}). For this experiment, we use the
feature-evaluation configuration (see \secref{subsec:setup}) and
compare \tool with and without the inline-assembly extension
enabled. Our metric is the percentage increase in instruction
coverage, measured as the share of instructions included in the
binaries generated from our product programs when the extension is
enabled.

Across the tested zkVMs, we observed consistent
improvements. Specifically, the inline-assembly extension increased
instruction coverage by 38.2\% in \risc, 15.0\% in \nexus, 45.2\% in
\jolt, 44.0\% in \spOne, 34.5\% in \openvm, and 44.0\% in \pico. These
gains demonstrate that the extension is effective at systematically
incorporating rarely used instructions that would otherwise remain
uncovered.

To provide deeper insight, \appref{sect:appendix:results} includes bar
charts showing the instruction-frequency distributions in binaries
generated from \tool programs, with and without the inline-assembly
extension.

\begin{figure*}[t]
  \centering
  \includegraphics[scale=0.45]{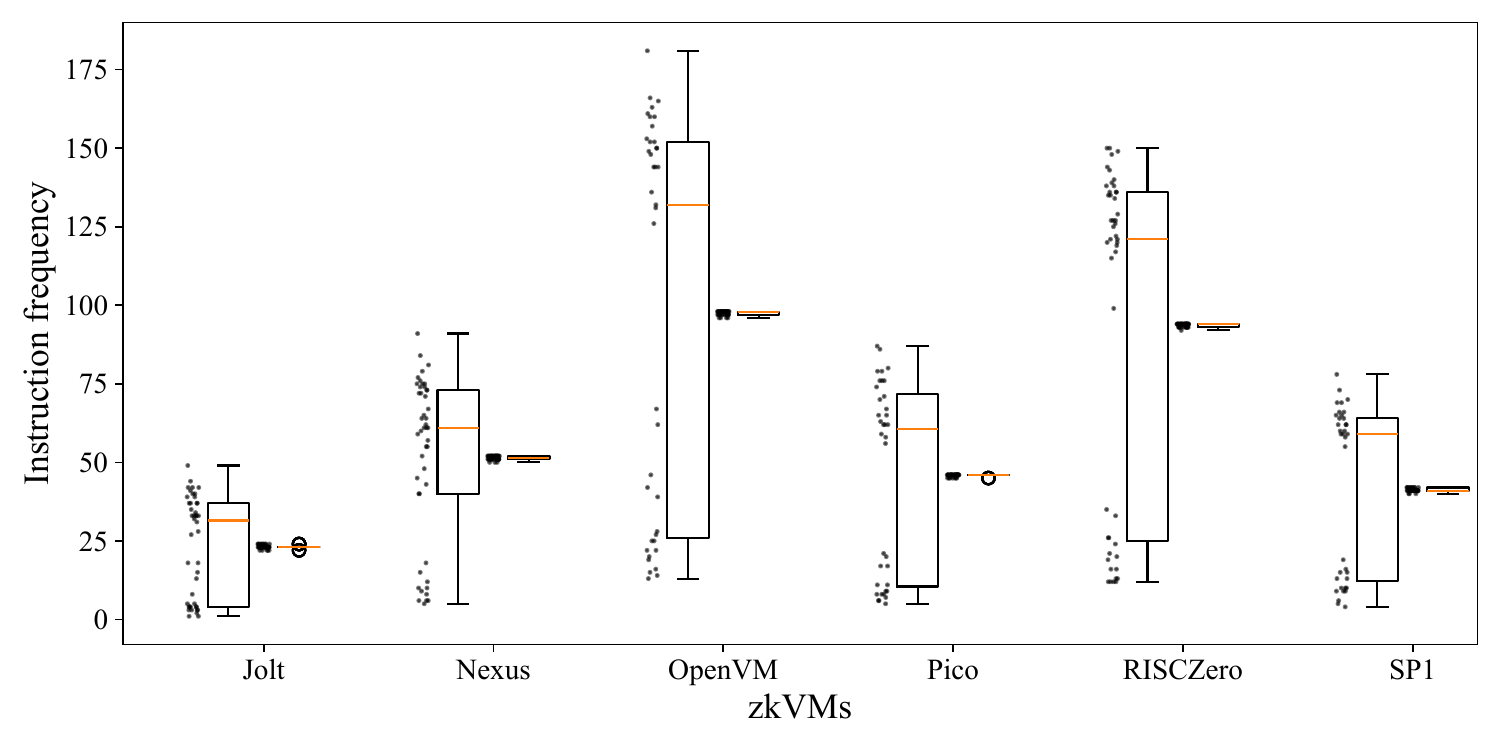}
  \caption{Distribution of injection frequencies across instructions
    for each zkVM, comparing \tool without (left box) and with (right
    box) the fault-injection scheduler.}
  \label{fig:rq5}
\end{figure*}

\paragraph{RQ5: Effectiveness of fault-injection scheduler.}
In this research question, we evaluate the effectiveness of the
fault-injection scheduler, whose goal is to apply injections uniformly
across all available \riscv instructions (see \secref{sec:step7}). For
this experiment, we use the feature-evaluation configuration (see
\secref{subsec:setup}) and compare \tool with and without the
scheduler enabled. Without the scheduler, injections are placed at
random points in the trace, which disproportionately targets common
instructions and leaves rare ones underexplored. With the scheduler,
injections are guided by instruction frequency to ensure more uniform
coverage.

\figref{fig:rq5} shows the results as box plots. The x-axis denotes
the zkVMs, while the y-axis shows how often each instruction was
selected for injection. For each zkVM, the left box represents \tool
without the scheduler, and the right box \tool with the scheduler. As
the figure illustrates, the scheduler prevents imbalance by
distributing injections evenly, thereby avoiding undertesting of rare
instructions.

\paragraph{RQ6: Impact of instruction-modification injection.}
Here, we measure the impact of the instruction-modification injection
(see \secref{sec:step7}), which is the only injection type implemented
across all zkVMs we tested and the one responsible for uncovering all
of our soundness bugs. For this experiment, we used the
feature-evaluation configuration (see \secref{subsec:setup}).

\tabref{tab:injection} summarizes the results. The first column lists
each zkVM, followed by the total number of injections applied. The
remaining columns categorize their effects:
\begin{itemize}
\item \textbf{\code{SUCCESS}, \code{EC == 0}:} The product program
  produced the expected output (i.e., \code{SUCCESS}), and the zkVM
  terminated successfully, that is, with exit code (EC) zero. So these
  injections had no observable effect---the left sub-column shows
  their absolute number and the right sub-column their percentage.
\item \textbf{\code{SUCCESS}, \code{EC != 0}:} The output of the
  product program was unchanged, but the zkVM terminated
  unsuccessfully (e.g., crashed).
\item \textbf{\code{OOPS}, \code{EC == 0}:} This case would correspond
  to a soundness bug (unexpected output while the verifier
  succeeds). As expected, no such cases occurred here.
\item \textbf{\code{OOPS}, \code{EC != 0}:} The output of the product
  program was altered, while the zkVM execution itself failed.
\end{itemize}

The last column is the most relevant: it shows that in the vast
majority of cases, the instruction-modification injection effectively
perturbs executions so that the program output changes. Such output
divergence is a crucial prerequisite for detecting soundness bugs
caused by underconstrained behavior (see \secref{sec:step7}).

\begin{table*}[t]
\centering
\begin{tabular}{c|c|r|r|r|r|r|r|r|r}
  \textbf{zkVM} &
  \textbf{\makecell{Total\\injections}} &
  \multicolumn{2}{c|}{\textbf{\makecell{\code{SUCCESS},\\ \code{EC == 0}}}} &
  \multicolumn{2}{c|}{\textbf{\makecell{\code{SUCCESS},\\ \code{EC != 0}}}} &
  \multicolumn{2}{c|}{\textbf{\makecell{\code{OOPS},   \\ \code{EC == 0}}}} &
  \multicolumn{2}{c}{\textbf{\makecell{\code{OOPS},    \\ \code{EC != 0}}}} \\
  \hline
  \risc   & 4404 & 275 &  6.2\% &   0 &  0.0\% & 0 & 0.0\% & 4129 &  93.8\% \\
  \nexus  & 2366 & 476 & 20.1\% &  83 &  3.5\% & 0 & 0.0\% & 1807 &  76.4\% \\
  \jolt   & 1040 &   2 &  0.2\% & 445 & 42.8\% & 0 & 0.0\% &  593 &  57.0\% \\
  \spOne  & 1483 &  32 &  2.2\% & 833 & 56.2\% & 0 & 0.0\% &  618 &  41.7\% \\
  \openvm & 3801 & 192 &  5.1\% & 386 & 10.2\% & 0 & 0.0\% & 3223 &  84.8\% \\
  \pico   & 1649 &   0 &  0.0\% &   0 &  0.0\% & 0 & 0.0\% & 1649 & 100.0\% \\
\end{tabular}
\caption{Results of the instruction-modification injection across all
  tested zkVMs. The table shows the total number of injections and
  their outcomes. \code{SUCCESS} means the product program output was
  correct, \code{OOPS} indicates an altered output, and \code{EC}
  denotes the exit code returned by the zkVM.}
\label{tab:injection}
\end{table*}

%%% Local Variables:
%%% mode: latex
%%% TeX-master: "main"
%%% End:

%!TEX root = main.tex

%-------------------------------------------------------------------------------
\section{Related Work}
\label{sec:related}
%-------------------------------------------------------------------------------

We present the first systematic fuzzing approach for testing zkVMs. It
combines metamorphic testing with fault injection to uncover both
soundness and completeness bugs. While prior work has examined testing
techniques for other classes of zero-knowledge systems, we are not
aware of any existing work that targets zkVMs end-to-end. Despite the
growing importance of zkVMs in blockchain rollups and
privacy-preserving computation, their correctness has so far remained
an open challenge in the literature.

In the remainder of this section, we focus on four main areas of
prior research relevant to our work: fuzzing of ZK pipelines,
metamorphic-testing approaches, fault-injection techniques, and
fuzzing of \riscv CPUs.

\paragraph{Fuzzing of ZK pipelines.}
The most closely related work to ours is
\circuzz~\cite{HochrainerIsychev2025}, which applies metamorphic
testing to detect bugs in ZK pipelines. While \circuzz targets systems
that take a domain-specific circuit language as input, \tool focuses
on zkVMs that execute general-purpose code, frequently in Rust, and
enforce constraints implicitly from the program semantics. This shift
in input language and execution model requires new techniques for
program generation and performance optimization.

MTZK~\cite{XiaoLiu2025} takes a narrower scope, focusing on testing
zero-knowledge compilers via metamorphic transformations. In contrast,
our work targets zkVMs (that typically rely on mature compilers, such
as the Rust compiler) and tests the full processing pipeline,
including execution, proof generation, and verification.

Finally, unlike either \circuzz or MTZK, \tool integrates a
fault-injection mechanism that can simulate a malicious prover. This
enables the detection of soundness bugs caused by overly weak
constraints, a class of vulnerabilities that metamorphic testing alone
cannot uncover.

\paragraph{Metamorphic testing.}
Metamorphic testing~\cite{ChenCheung1998} is widely used in domains
where a reliable test oracle is
unavailable~\cite{BarrHarman2015}. Segura et
al.~\cite{SeguraFraser2016} provide a comprehensive survey of its
applications across different domains.

The most closely related applications to our work are in the testing
of compilers~\cite{ChenPatra2020} and program analyzers~(e.g.,
\cite{ZhangSu2019,MansurChristakis2020,WintererZhang2020-Fusion,MansurChristakis2021,MansurWuestholz2023,ZhangPei2023,MordahlZhang2023,ZhangPei2024,HeDi2024,KaindlstorferIsychev2024}),
such as software model checkers~\cite{BiereCimatti1999,McMillan2018}
and abstract interpreters~\cite{CousotCousot1977}.
In these contexts, metamorphic testing typically generates two
programs that are syntactically different yet semantically equivalent,
and checks whether their outputs match.

Our work differs in that we embed both executions into a single
product program, rather than running them separately and comparing
results externally. This design enables knowing the correct output of
the product program in advance and often reduces execution overhead,
especially in the prover. While metamorphic testing provides a
natural way to design oracles, our work complements it with fault
injection, which explores a different dimension of zkVM robustness.

\paragraph{Fault injection.}
Fault injection~\cite{ArlatAguera1990,ClarkPradhan1995,HsuehTsai1997}
has long been used to evaluate the robustness and
dependability of systems.
Building on these foundations, practical frameworks such as
LFI~\cite{MarinescuCandea2009} and dynamic stub
injection~\cite{ChristakisEmmisberger2017} provide general-purpose
mechanisms. These tools allow developers to inject faults at the
library or API boundary to test error-handling code, independent of
the underlying application logic.

More recent work combines fuzzing and fault injection. For example,
\textsc{Fuzztruction}~\cite{BarsSchloegel2023} and
\textsc{Fuzztruction-Net}~\cite{BarsSchloegel2024} explore
cross-application testing by generating and consuming data across
paired applications, such as encryption or compression
tools. \textsc{FuzzERR}~\cite{SharmaTanksalkar2024} focuses on
inserting faults into API calls to evaluate robustness of error
handling, while context-sensitive software fault injection has been
used to fuzz error-handling code by tailoring injected faults to the
surrounding program
context~\cite{JiangBai2020}. IFIZZ~\cite{LiuJi2021} focuses on IoT
firmware, efficiently generating deep-state fault scenarios to test
resilience in resource-constrained environments.

In contrast, our work is the first to apply fault injection to zkVMs
by injecting faults directly into the VM's execution logic. This
simulates malicious prover behavior and tests whether the verifier can
be deceived---revealing soundness bugs specific to zkVM constraint
systems that are not addressed by prior fault-injection frameworks.

Although our approach does not infer what faults to inject, related
work has explored automatically inferring likely faults and error
specifications~\cite{MarinescuCandea2009,AcharyaXie2009,JanaKang2016}
in other domains.

\paragraph{Fuzzing of \riscv CPUs.}
There are fuzzers that have targeted \riscv CPUs, such as
\cite{SoltCeesaySeitz2024,XuLiu2023,KandeCrump2022,HurSong2021},
aiming to check if the hardware correctly implements the \riscv
instruction set architecture.
Our work is fundamentally different and complementary to these
hardware-focused approaches as \tool operates entirely at the software
layer. Of course, a zkVM needs both a correctly functioning hardware
CPU to run on (which CPU fuzzers like the above test) and correct
software logic (which \tool validates).

%%% Local Variables:
%%% mode: latex
%%% TeX-master: "main"
%%% End:

%!TEX root = main.tex

%-------------------------------------------------------------------------------
\section{Conclusion}
\label{sec:conclusion}
%-------------------------------------------------------------------------------

We presented \tool, the first automated fuzzer for detecting soundness
and completeness bugs in zkVMs. Our approach introduces a novel
testing methodology that combines an efficient, product-program-based
variant of metamorphic testing with a fault-injection mechanism
designed to simulate malicious provers.
Our evaluation demonstrates that \tool is effective across multiple
zkVM implementations, uncovering both soundness and completeness
bugs. The modular design allows it to be easily adapted to other zkVMs
with modest engineering effort, positioning it as a practical tool for
improving the reliability of this rapidly evolving ecosystem.

For future work, we plan to extend \tool in several directions. First,
we aim to enhance the circuit generator to produce more complex and
stateful programs, which could uncover deeper bugs in the VM execution
logic. Second, exploring more sophisticated fault-injection
strategies
may uncover deeper vulnerabilities. Finally, we plan to adapt \tool to
support zkVMs based on instruction sets beyond \riscv, further
broadening its impact and helping to secure a wider range of such
systems.

%%% Local Variables:
%%% mode: latex
%%% TeX-master: "main"
%%% End:

%% %-------------------------------------------------------------------------------
%% \section*{Acknowledgments}
%% %-------------------------------------------------------------------------------

%% The USENIX latex style is old and very tired, which is why
%% there's no \textbackslash{}acks command for you to use when
%% acknowledging. Sorry.

%% \textbf{Do not include any acknowledgements in your submission which may deanonymize you (e.g., because of specific affiliations or grants you acknowledge)}

%-------------------------------------------------------------------------------
% optional clearing of the page
\cleardoublepage
\appendix

%!TEX root = main.tex

\onecolumn

%%----------------------------------------------------------------------------
\section{\tool Rewrite Rules}
\label{sect:appendix:rules}
%%----------------------------------------------------------------------------

\newcolumntype{Y}{>{\centering\arraybackslash}X}

\begin{table}[h!]
    \centering
    \small
    \begin{adjustbox}{center}
    \begin{tabularx}{\textwidth}{*{3}{Y}}
    \toprule
    {\normalsize\textbf{Rule ID}} & {\normalsize\textbf{Match Pattern}} & {\normalsize\textbf{Rewrite Template}} \\
    \midrule
    \texttt{comm-or}                 & \texttt{?a | ?b}               & \texttt{?b | ?a} \\
    \texttt{assoc-and}               & \texttt{(?a \& ?b) \& ?c}      & \texttt{?a \& (?b \& ?c)} \\
    \texttt{comm-and}                & \texttt{?a \& ?b}              & \texttt{?b \& ?a} \\
    \texttt{and-zero}                & \texttt{?a \& 0}               & \texttt{0} \\
    \texttt{inv-xor}                 & \texttt{?a \^{} ?a}            & \texttt{0} \\
    \texttt{comm-xor}                & \texttt{?a \^{} ?b}            & \texttt{?b \^{} ?a} \\
    \texttt{zero-or-rev}             & \texttt{?a | 0}                & \texttt{?a} \\
    \texttt{zero-xor-rev}            & \texttt{?a \^{} 0}             & \texttt{?a} \\
    \texttt{inv-xor-rev}             & \texttt{0}                     & \texttt{(\$r:int \^{} \$r:int)} \\
    \texttt{zero-or}                 & \texttt{?a:int}                & \texttt{(?a | 0)} \\
    \texttt{zero-xor}                & \texttt{?a:int}                & \texttt{(?a \^{} 0)} \\
    \texttt{idem-and}                & \texttt{?a:int}                & \texttt{(?a \& ?a)} \\
    \texttt{zero-and}                & \texttt{0}                     & \texttt{(\$r:int \& 0)} \\
    \texttt{one-div}                 & \texttt{1}                     & \texttt{(\$r:int / \$r:int)} \\
    \texttt{comm-add}                & \texttt{?a + ?b}               & \texttt{?b + ?a} \\
    \texttt{comm-mul}                & \texttt{?a * ?b}               & \texttt{?b * ?a} \\
    \texttt{dist-mul-add}            & \texttt{(?a + ?b) * ?c}        & \texttt{(?a * ?c) + (?b * ?c)} \\
    \texttt{dist-add-mul}            & \texttt{(?a * ?c) + (?b * ?c)} & \texttt{(?a + ?b) * ?c} \\
    \texttt{assoc-add}               & \texttt{(?a + ?b) + ?c}        & \texttt{?a + (?b + ?c)} \\
    \texttt{assoc-add-rev}           & \texttt{?a + (?b + ?c)}        & \texttt{(?a + ?b) + ?c} \\
    \texttt{assoc-mul}               & \texttt{(?a * ?b) * ?c}        & \texttt{?a * (?b * ?c)} \\
    \texttt{assoc-mul-rev}           & \texttt{?a * (?b * ?c)}        & \texttt{(?a * ?b) * ?c} \\
    \texttt{zero-add-des}            & \texttt{?a + 0}                & \texttt{?a} \\
    \texttt{one-mul-des}             & \texttt{?a * 1}                & \texttt{?a} \\
    \texttt{one-div-des}             & \texttt{?a / 1}                & \texttt{?a} \\
    \texttt{inv-zero-add-des}        & \texttt{?a - 0}                & \texttt{?a} \\
    \texttt{inv-add-des}             & \texttt{?a - ?a}               & \texttt{0} \\
    \texttt{inv-assoc-neg2pos}       & \texttt{(?a - ?b) - ?c}        & \texttt{?a - (?b + ?c)} \\
    \texttt{inv-assoc-pos2neg}       & \texttt{?a - (?b + ?c)}        & \texttt{(?a - ?b) - ?c} \\
    \texttt{pow2-to-mul}             & \texttt{?a ** 2}               & \texttt{?a * ?a} \\
    \texttt{pow3-to-mul}             & \texttt{?a ** 3}               & \texttt{(?a * ?a) * ?a} \\
    \texttt{mul-to-pow2}             & \texttt{?a * ?a}               & \texttt{?a ** 2} \\
    \texttt{mul-to-pow3}             & \texttt{(?a * ?a) * ?a}        & \texttt{?a ** 3} \\
    \texttt{zero-add-con}            & \texttt{?a:int}                & \texttt{?a + 0} \\
    \texttt{one-mul-con}             & \texttt{?a:int}                & \texttt{?a * 1} \\
    \texttt{one-div-con}             & \texttt{?a:int}                & \texttt{?a / 1} \\
    \texttt{rem-of-one-con}          & \texttt{0}                     & \texttt{\$r:int \% 1} \\
    \texttt{rem-of-one-des}          & \texttt{?a \% 1}               & \texttt{0} \\
    \texttt{and-to-rem}              & \texttt{?a \& 1}               & \texttt{?a \% 2} \\
    \texttt{rem-to-and}              & \texttt{?a \% 2}               & \texttt{?a \& 1} \\
    \texttt{inv-zero-add-con}        & \texttt{?a:int}                & \texttt{?a - 0} \\
    \texttt{inv-addition-exp}        & \texttt{?a - ?c}               & \texttt{?a + (0 - ?c)} \\
    \texttt{double-negation-add-con} & \texttt{?a:int}                & \texttt{0 - (0 - ?a)} \\
    \texttt{add-sub-random-value}    & \texttt{?a:int}                & \texttt{(?a - \$r:int) + \$r:int} \\
    \texttt{zero-lor-des}            & \texttt{?a || F}               & \texttt{?a} \\
    \texttt{zero-land-des}           & \texttt{?a \&\& T}             & \texttt{?a} \\
    \texttt{taut-lor}                & \texttt{?a || T}               & \texttt{T} \\
    \texttt{contra-land}             & \texttt{?a \&\& F}             & \texttt{F} \\
    \texttt{assoc-lor}               & \texttt{(?a || ?b) || ?c}                   & \texttt{?a || (?b || ?c)} \\
    \texttt{assoc-land}              & \texttt{(?a \&\& ?b) \&\& ?c}               & \texttt{?a \&\& (?b \&\& ?c)} \\    
    \texttt{comm-lor}                & \texttt{?a || ?b}                           & \texttt{?b || ?a} \\
    \texttt{comm-lan}                & \texttt{?a \&\& ?b}                         & \texttt{?b \&\& ?a} \\
    \texttt{dist-lor-land}           & \texttt{(?a \&\& ?b) || ?c}                 & \texttt{(?a || ?c) \&\& (?b || ?c)} \\
    \texttt{dist-land-lor}           & \texttt{(?a || ?c) \&\& (?b || ?c)}         & \texttt{(?a \&\& ?b) || ?c} \\    
    \bottomrule
    \end{tabularx}
    \end{adjustbox}
\end{table}

\newpage

\begin{table}[t!]
    \centering
    \small  
    \begin{adjustbox}{center}
    \begin{tabularx}{\textwidth}{*{3}{Y}}
    \toprule
    {\normalsize\textbf{Rule ID}} & {\normalsize\textbf{Match Pattern}} & {\normalsize\textbf{Rewrite Template}} \\  
    \midrule
    \texttt{de-morgan-land-con}          & \texttt{!(?a \&\& ?b)}                      & \texttt{(!?a) || (!?b)} \\
    \texttt{de-morgan-land-des}          & \texttt{(!?a) || (!?b)}                     & \texttt{!(?a \&\& ?b)} \\
    \texttt{de-morgan-lor-con}           & \texttt{!(?a || ?b)}                        & \texttt{(!?a) \&\& (!?b)} \\
    \texttt{de-morgan-lor-des}           & \texttt{(!?a) \&\& (!?b)}                   & \texttt{!(?a || ?b)} \\
    \texttt{double-negation-des}         & \texttt{!(!?a)}                             & \texttt{?a} \\
    \texttt{double-land-des}             & \texttt{?a \&\& ?a}                         & \texttt{?a} \\
    \texttt{double-lor-des}              & \texttt{?a || ?a}                           & \texttt{?a} \\
    \texttt{double-lxor-des}             & \texttt{?a \^{}\^{} ?a}                     & \texttt{F} \\
    \texttt{comm-lxor}                   & \texttt{?a \^{}\^{} ?b}                     & \texttt{?b \^{}\^{} ?a} \\
    \texttt{lxor-to-or-and}              & \texttt{?a \^{}\^{} ?b}                     & \texttt{((!?a) \&\& ?b) || (?a \&\& (!?b))} \\
    \texttt{zero-lor-con}                & \texttt{?a:bool}                            & \texttt{?a || F} \\
    \texttt{zero-land-con}               & \texttt{?a:bool}                            & \texttt{?a \&\& T} \\
    \texttt{double-negation-con}         & \texttt{?a:bool}                            & \texttt{!(!?a)} \\
    \texttt{double-land-con}             & \texttt{?a:bool}                            & \texttt{?a \&\& ?a} \\
    \texttt{double-lor-con}              & \texttt{?a:bool}                            & \texttt{?a || ?a} \\
    \texttt{double-lxor-con}             & \texttt{F}                                  & \texttt{\$r:bool \^{}\^{} \$r:bool} \\
    \texttt{or-and-to-lxor}              & \texttt{((!?a) \&\& ?b) || (?a \&\& (!?b))} & \texttt{?a \^{}\^{} ?b} \\
    \texttt{commutativity-equ}           & \texttt{?a == ?b}                           & \texttt{?b == ?a} \\
    \texttt{relation-geq-to-leq}         & \texttt{?a >= ?b}                           & \texttt{?b <= ?a} \\
    \texttt{relation-leq-to-geq}         & \texttt{?a <= ?b}                           & \texttt{?b >= ?a} \\
    \texttt{relation-leq-to-lth-and-equ} & \texttt{?a <= ?b}                           & \texttt{(?a < ?b) || (?a == ?b)} \\
    \texttt{relation-lth-and-equ-to-leq} & \texttt{(?a < ?b) || (?a == ?b)}            & \texttt{?a <= ?b} \\
    \texttt{relation-geq-to-gth-and-equ} & \texttt{?a >= ?b}                           & \texttt{(?a > ?b) || (?a == ?b)} \\
    \texttt{relation-gth-and-equ-to-geq} & \texttt{(?a > ?b) || (?a == ?b)}            & \texttt{?a >= ?b} \\
    \texttt{relation-leq-to-not-gth}     & \texttt{?a <= ?b}                           & \texttt{!(?a > ?b)} \\
    \texttt{relation-not-gth-to-leq}     & \texttt{!(?a > ?b)}                         & \texttt{?a <= ?b} \\
    \texttt{relation-geq-to-not-lth}     & \texttt{?a >= ?b}                           & \texttt{!(?a < ?b)} \\
    \texttt{relation-not-lth-to-geq}     & \texttt{!(?a < ?b)}                         & \texttt{?a >= ?b} \\
    \texttt{relation-neq-to-not-equ}     & \texttt{?a != ?b}                           & \texttt{!(?a == ?b)} \\
    \texttt{relation-not-equ-to-neq}     & \texttt{!(?a == ?b)}                        & \texttt{?a != ?b} \\
    \bottomrule
    \end{tabularx}
    \end{adjustbox}
\end{table}

\todo{}

\twocolumn

%%% Local Variables:
%%% mode: latex
%%% TeX-master: "main"
%%% End:

\cleardoublepage
%!TEX root = main.tex

%%----------------------------------------------------------------------------
\section{\tool Injection Types}
\label{sect:appendix:injections}
%%----------------------------------------------------------------------------

\paragraph{Injection types per zkVM.}
We list below the injection types that \tool supports for each zkVM implementation.

\bigskip
\medskip

\noindent
\textbf{\jolt}
\begin{itemize}
\item \code{INSTR_WORD_MOD}
\end{itemize}

\noindent
\textbf{\nexus}
\begin{itemize}
\item \code{INSTR_WORD_MOD}
\item \code{POST_EXEC_PC_MOD}
\end{itemize}

\noindent
\textbf{\openvm}
\begin{itemize}
\item \code{INSTR_WORD_MOD}
\item \code{BASE_ALU_RANDOM_OUTPUT}
\item \code{LOADSTORE_SHIFT_MOD}
\item \code{LOADSTORE_OPCODE_MOD}
\item \code{LOADSTORE_SKIP_WRITE}
\item \code{LOADSTORE_PC_MOD}
\item \code{LOAD_SIGN_EXTEND_SHIFT_MOD}
\item \code{LOAD_SIGN_EXTEND_MSB_FLIPPED}
\item \code{LOAD_SIGN_EXTEND_MSL_FLIPPED}
\item \code{DIVREM_FLIP_IS_SIGNED}
\item \code{DIVREM_FLIP_IS_DIV}
\item \code{AUIPC_PC_LIMBS_MODIFICATION}
\item \code{AUIPC_IMM_LIMBS_MODIFICATION}
\end{itemize}

\noindent
\textbf{\pico}
\begin{itemize}
\item \code{INSTR_WORD_MOD}
\item \code{EMULATE_RANDOM_INSTRUCTION}
\item \code{MODIFY_OUTPUT_VALUE}
\end{itemize}

\noindent
\textbf{\risc}
\begin{itemize}
\item \code{PRE_EXEC_PC_MOD}
\item \code{POST_EXEC_PC_MOD}
\item \code{INSTR_WORD_MOD}
\item \code{BR_NEG_COND}
\item \code{COMP_OUT_MOD}
\item \code{LOAD_VAL_MOD}
\item \code{STORE_OUT_MOD}
\item \code{PRE_EXEC_MEM_MOD}
\item \code{POST_EXEC_MEM_MOD}
\item \code{PRE_EXEC_REG_MOD}
\item \code{POST_EXEC_REG_MOD}
\end{itemize}

\noindent
\textbf{\spOne}
\begin{itemize}
\item \code{POST_EXEC_PRE_COMMIT_PC_MOD}
\item \code{POST_EXEC_POST_COMMIT_PC_MOD}
\item \code{INSTR_WORD_MOD}
\item \code{ALU_RESULT_MOD}
\item \code{ALU_RESULT_LOC_MOD}
\item \code{ALU_PARSED_OPERAND_MOD}
\item \code{EXECUTE_INSTRUCTION_AGAIN}
\item \code{ALU_LOAD_OPERAND_MOD}
\item \code{SYS_CALL_MOD_ECALL_ID}
\end{itemize}

\paragraph{Injection-type descriptions.}
In the following, we provide a brief overview of the main injection
types introduced above. A group of injections directly target
instruction execution.  \code{INSTR_WORD_MOD} modifies an instruction
before it is executed---this corresponds to the
instruction-modification injection presented in
\secref{sec:overview}. \code{EMULATE_RANDOM_INSTRUCTION} inserts and
executes a random instruction, while \code{EXECUTE_INSTRUCTION_AGAIN}
simply re-executes the same instruction.

A second group focuses on manipulating the program counter under
different circumstances. \code{PRE_EXEC_PC_MOD},
\code{POST_EXEC_PC_MOD}, \code{POST_EXEC_PRE_COMMIT_PC_MOD}, and
\code{POST_EXEC_POST_COMMIT_PC_MOD} modify the program counter, but at
distinct points in the execution flow---before or after executing an
instruction, or before or after committing the instruction to the
trace record.

Another group of injections targets operands and output values. For
example, \code{ALU_LOAD_OPERAND_MOD} and \code{ALU_PARSED_OPERAND_MOD}
modify instruction operands during and after parsing.
\code{BASE_ALU_RANDOM_OUTPUT}, \code{ALU_RESULT_MOD}, and
\code{MODIFY_OUTPUT_VALUE} alter the result of an operation before it
is written to a register or memory, while \code{COMP_OUT_MOD}
specifically modifies the output of side-effect-free instructions. In
\risc, we also support memory-specific variants: \code{LOAD_VAL_MOD},
which changes the value loaded from memory, and \code{STORE_OUT_MOD},
which modifies the value written to memory.

Next, we consider injections that target individual
instructions. Examples include \code{DIVREM_FLIP_IS_SIGNED} and
\code{AUIPC_PC_LIMBS_MODIFICATION}. These were designed for \openvm,
whose chip-based architecture facilitates fine-grained targeting of
particular instruction families. Other examples are
\code{SYS_CALL_MOD_ECALL_ID}, which alters the identifier of a system
call and thereby changes which system call is executed, and
\code{BR_NEG_COND}, which inverts a branch condition to flip the
direction of control flow.

Finally, we explored injections that tamper with where output values
are written or directly modify state. For example,
\code{ALU_RESULT_LOC_MOD} redirects the output of an operation to a
different destination. Similarly, \code{PRE_EXEC_MEM_MOD} and
\code{POST_EXEC_MEM_MOD} write arbitrary values into memory, while
\code{PRE_EXEC_REG_MOD} and \code{POST_EXEC_REG_MOD} do the same for
registers.

%%% Local Variables:
%%% mode: latex
%%% TeX-master: "main"
%%% End:

\cleardoublepage
%!TEX root = main.tex

%%----------------------------------------------------------------------------
\section{Additional Experimental Results}
\label{sect:appendix:results}
%%----------------------------------------------------------------------------

In this appendix, we provide additional results to support our
experimental evaluation (see \secref{sec:evaluation}).

\tabref{tab:refinding-minmax} summarizes the detailed results of the
bug-refinding experiment from RQ3. The first column lists the bug ID
(as defined in \tabref{tab:bugs}). The second column shows the number
of random seeds (out of five) for which the bug was rediscovered. The
remaining columns report the minimum, maximum, and median time to
rediscovery, as well as the corresponding minimum, maximum, and median
number of tested product programs until the bug was found.

\begin{table*}[t!]
\centering
\begin{tabular}{c|c|c|c|c|c|c|c|c|c}
  \textbf{Bug} &
  \textbf{Random} &
  \multicolumn{3}{c|}{\textbf{Time to bug}} &
  \multicolumn{3}{c|}{\textbf{Programs to bug}}\\
  \textbf{ID} &
  \textbf{seeds} &
  \textbf{min} &
  \textbf{max} &
  \textbf{median} &
  \textbf{min} &
  \textbf{max} &
  \textbf{median} \\
  \hline
  \bug{1}  & 5 / 5 & 00h35m16s & 08h06m57s & 04h33m59s &  20 & 484 & 259 \\
  \bug{2}  & 0 / 5 & -         & -         & -         & -   & -   & -   \\
  \bug{3}  & 2 / 5 & 07h58m29s & 17h20m41s & 12h39m35s & 200 & 440 & 320 \\
  \bug{4}  & 0 / 5 & -         & -         & -         & -   & -   & -   \\
  \bug{5}  & 2 / 5 & 01h59m42s & 08h17m16s & 05h08m29s &  52 & 215 & 133 \\
  \bug{6}  & 1 / 5 & 01h06m27s & 01h06m27s & 01h06m27s &   7 &   7 &   7 \\
  \bug{7}  & 0 / 5 & -         & -         & -         & -   & -   & -   \\
  \bug{8}  & 0 / 5 & -         & -         & -         & -   & -   & -   \\
  \bug{9}  & 5 / 5 & 00h02m44s & 00h33m57s & 00h19m30s &   1 &   9 &   5 \\
  \bug{10} & 5 / 5 & 00h02m54s & 00h31m45s & 00h08m06s &   1 &   7 &   2 \\
\end{tabular}
\caption{Detailed results of the bug-refinding experiment. For each
  bug ID, the table shows how many seeds rediscovered it, together
  with the min/median/max time and number of tested product programs
  that were required.}
\label{tab:refinding-minmax}
\end{table*}

\begin{figure*}[t]
  \centering
  \includegraphics[scale=0.55]{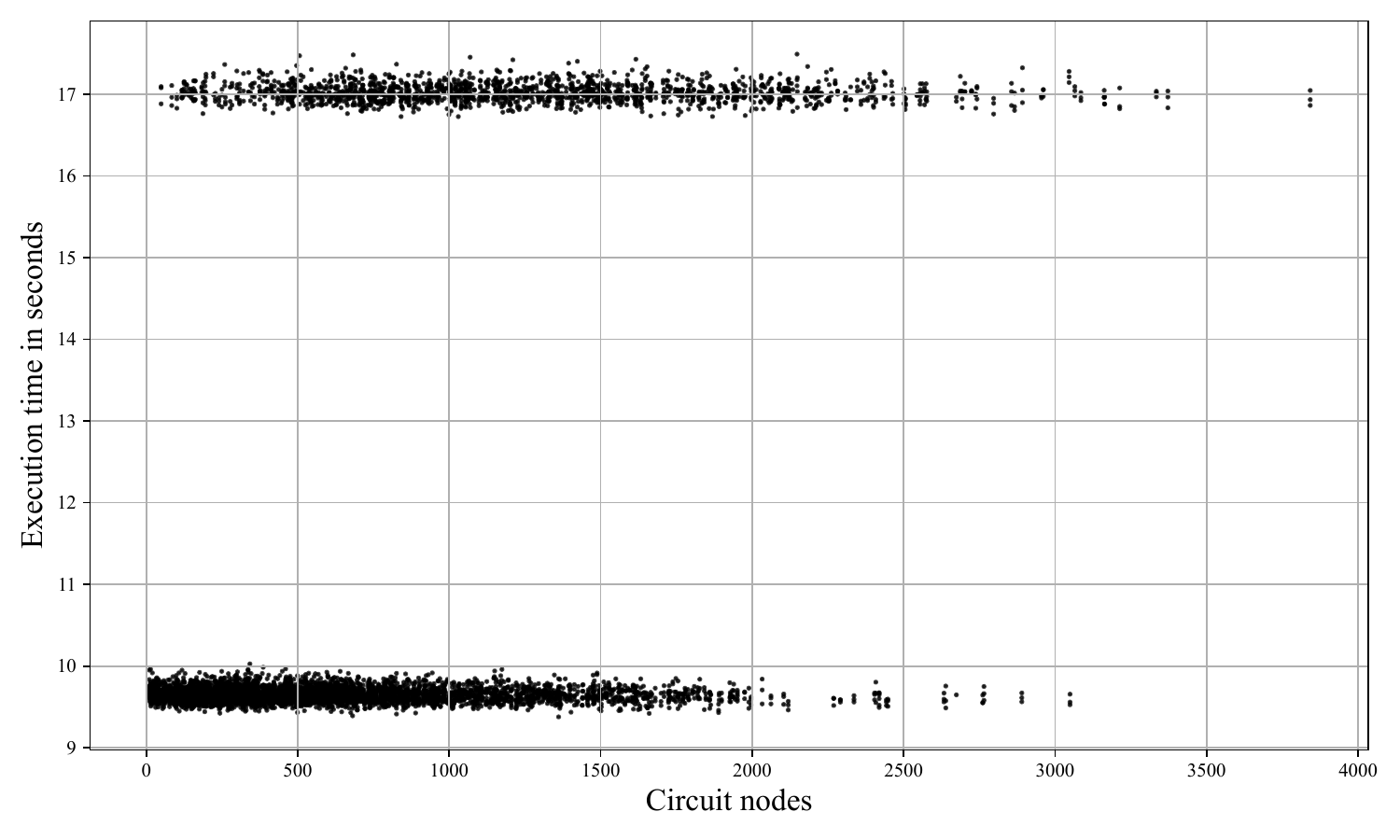}
  \caption{Execution time versus product-program size for \risc.}
  \label{fig:rq3-risc}
\end{figure*}

\begin{figure*}[t]
  \centering
  \includegraphics[scale=0.55]{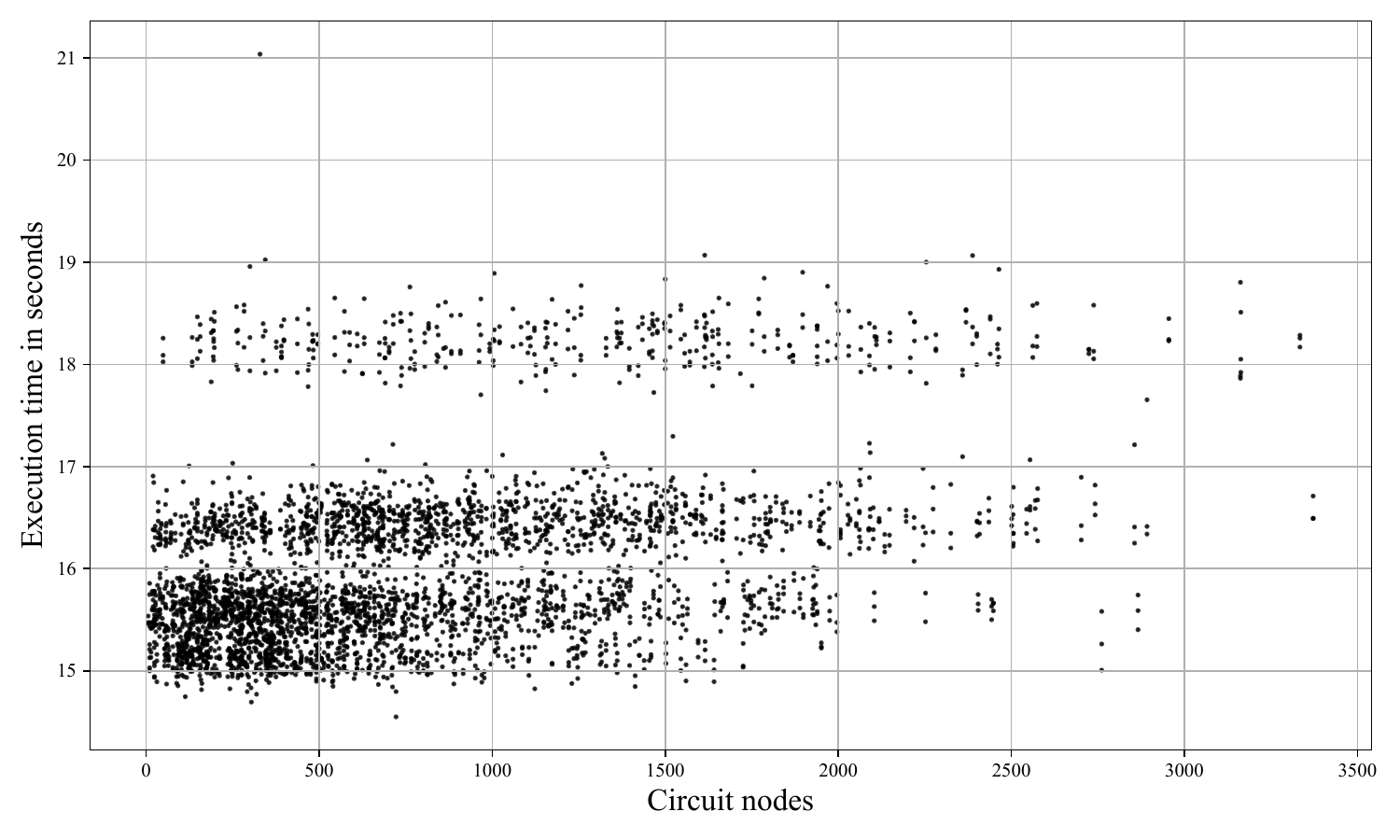}
  \caption{Execution time versus product-program size for \nexus.}
  \label{fig:rq3-nexus}
\end{figure*}

\begin{figure*}[t]
  \centering
  \includegraphics[scale=0.55]{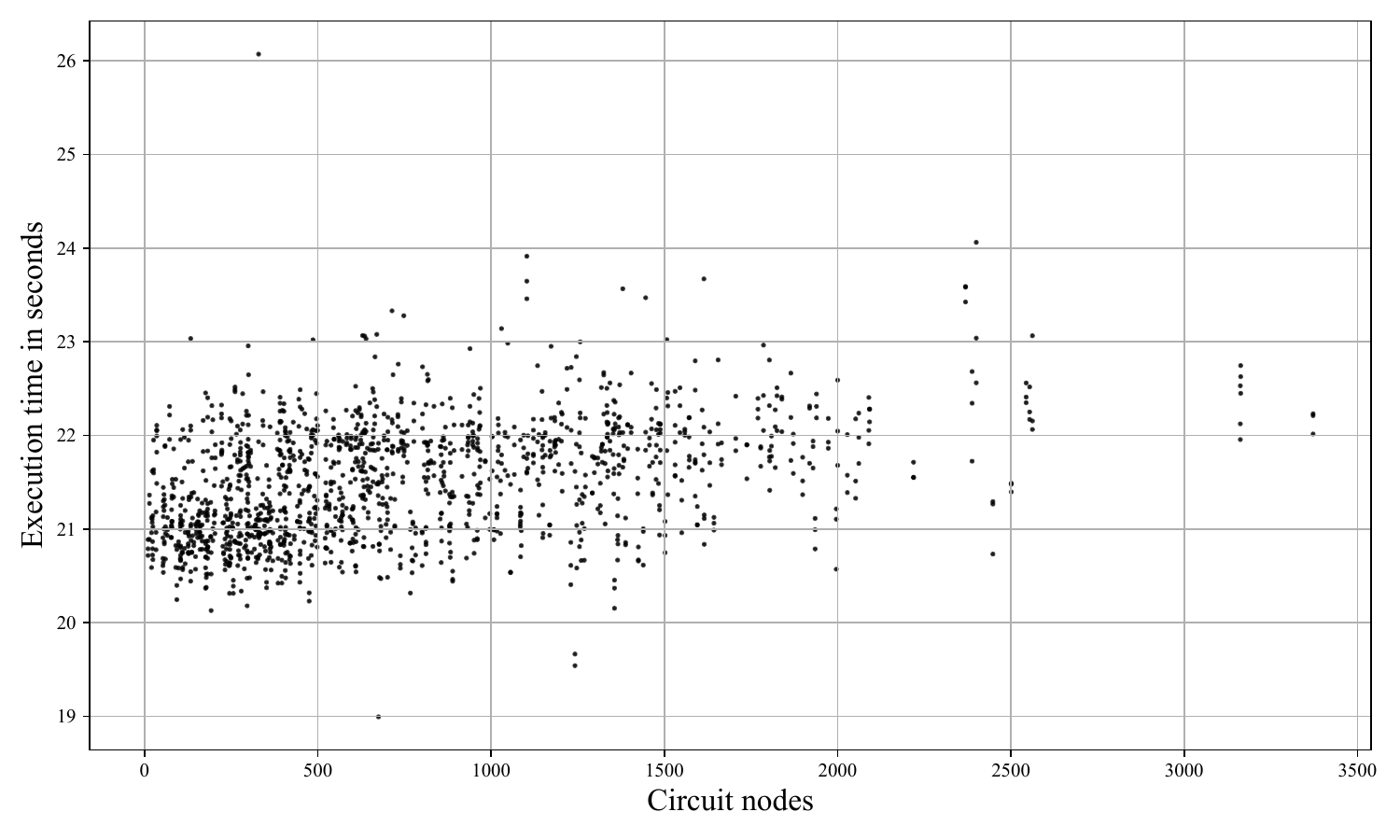}
  \caption{Execution time versus product-program size for \jolt.}
  \label{fig:rq3-jolt}
\end{figure*}

\begin{figure*}[t]
  \centering
  \includegraphics[scale=0.55]{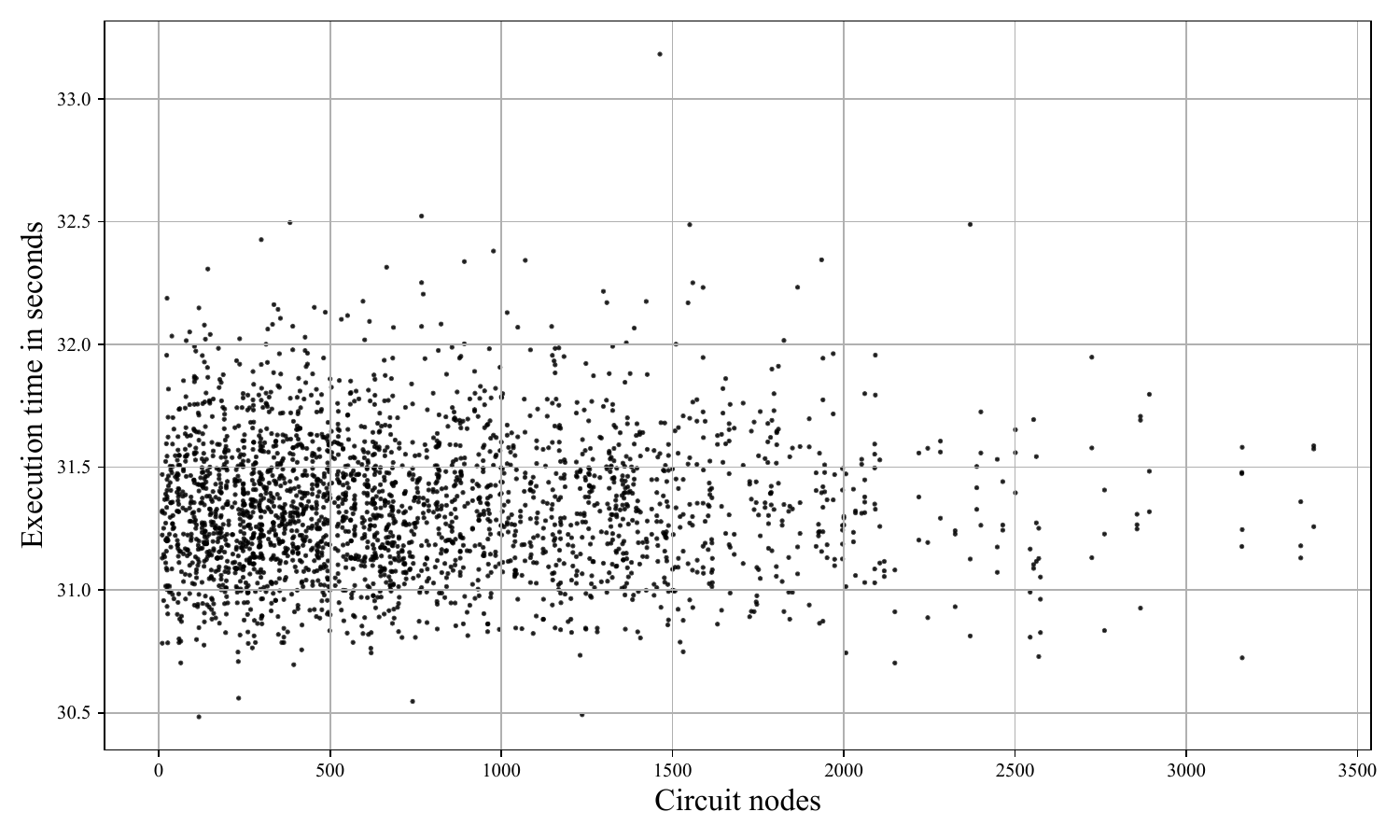}
  \caption{Execution time versus product-program size for \spOne.}
  \label{fig:rq3-sp1}
\end{figure*}

\begin{figure*}[t]
  \centering
  \includegraphics[scale=0.55]{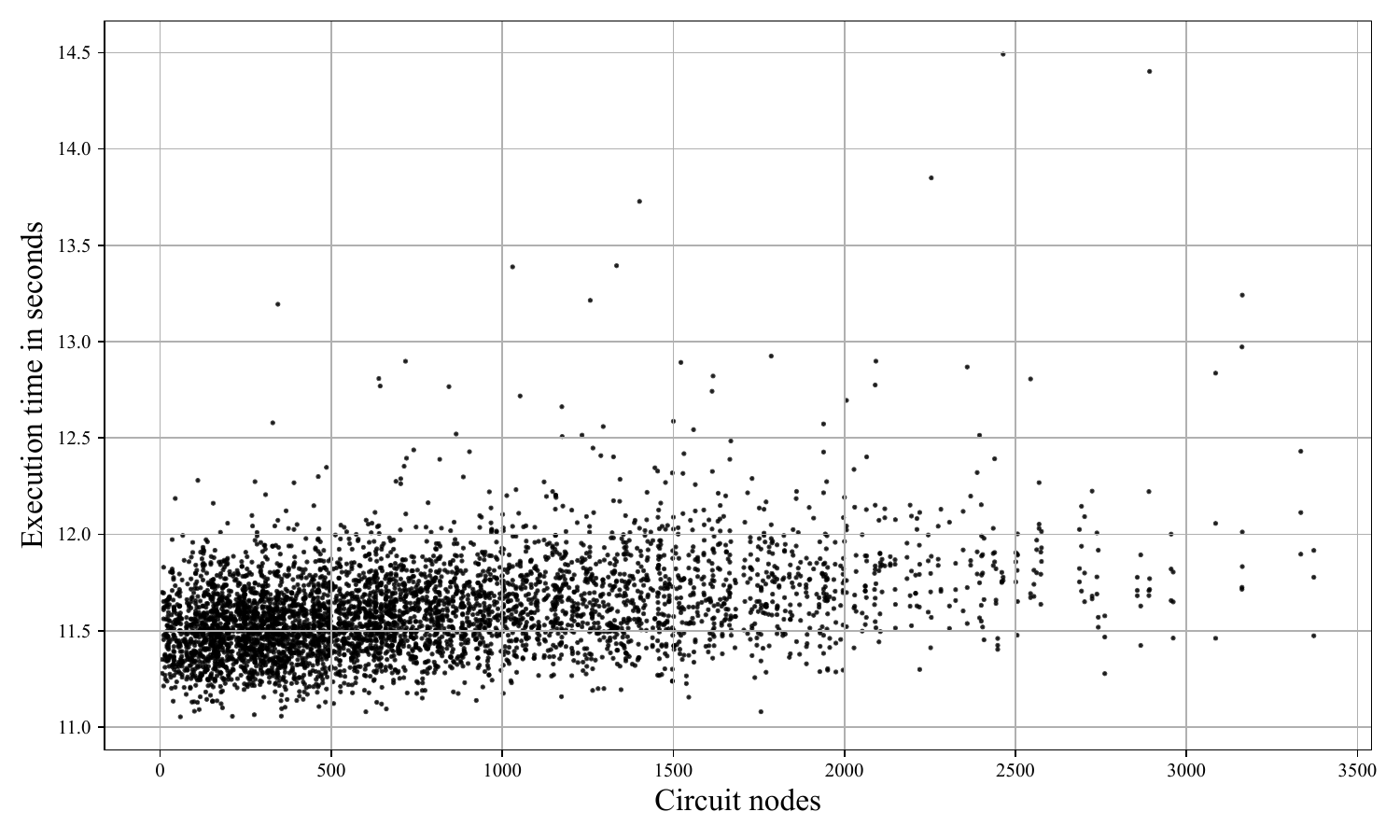}
  \caption{Execution time versus product-program size for \openvm.}
  \label{fig:rq3-openvm}
\end{figure*}

\begin{figure*}[t]
  \centering
  \includegraphics[scale=0.55]{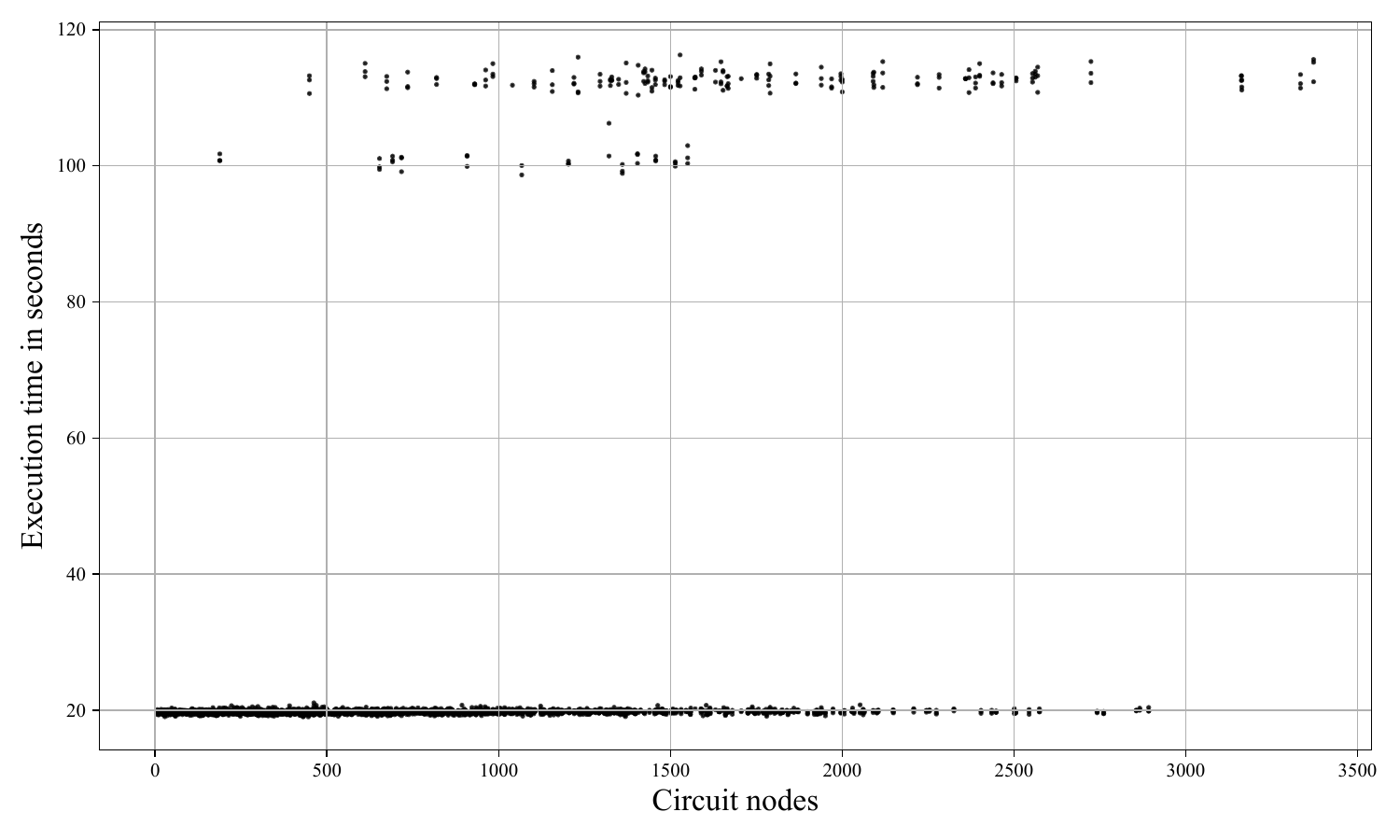}
  \caption{Execution time versus product-program size for \pico.}
  \label{fig:rq3-pico}
\end{figure*}

\figsref{fig:rq3-risc}--\ref{fig:rq3-pico} present scatter plots
illustrating the impact of product-program size on VM execution time
for each zkVM (see RQ3). The x-axis shows the product-program size, measured as
the number of AST nodes in the \il circuits used to generate the
program, and the y-axis shows the observed execution time.
For three VMs (i.e., \jolt, \spOne, and \openvm), we observe that
execution times fall within a very narrow range despite large
variation in program size. This confirms that execution time is not
strongly correlated with program size. For the other three VMs, we
observe distinct clusters of execution times. Programs in these
clusters differ in terms of their padded-trace sizes, and the jump in
runtime between clusters is explained by padding to the next power of
two. This is particularly obvious for \risc and \pico.

\begin{figure*}[t]
  \begin{subfigure}[b]{\textwidth}
    \centering
    \includegraphics[scale=0.55]{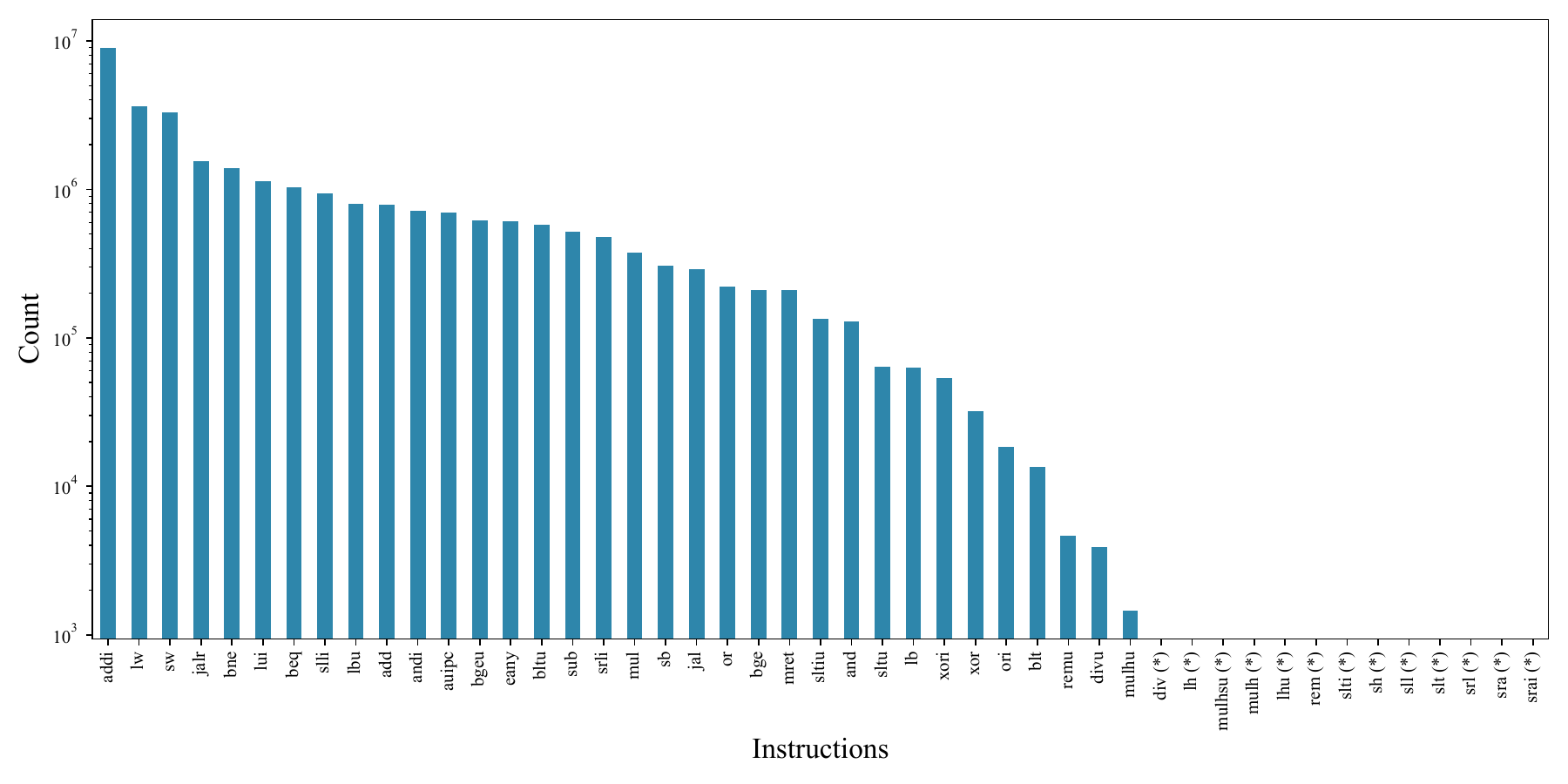}
  \end{subfigure}\\
  \begin{subfigure}[b]{\textwidth}
    \centering
    \includegraphics[scale=0.55]{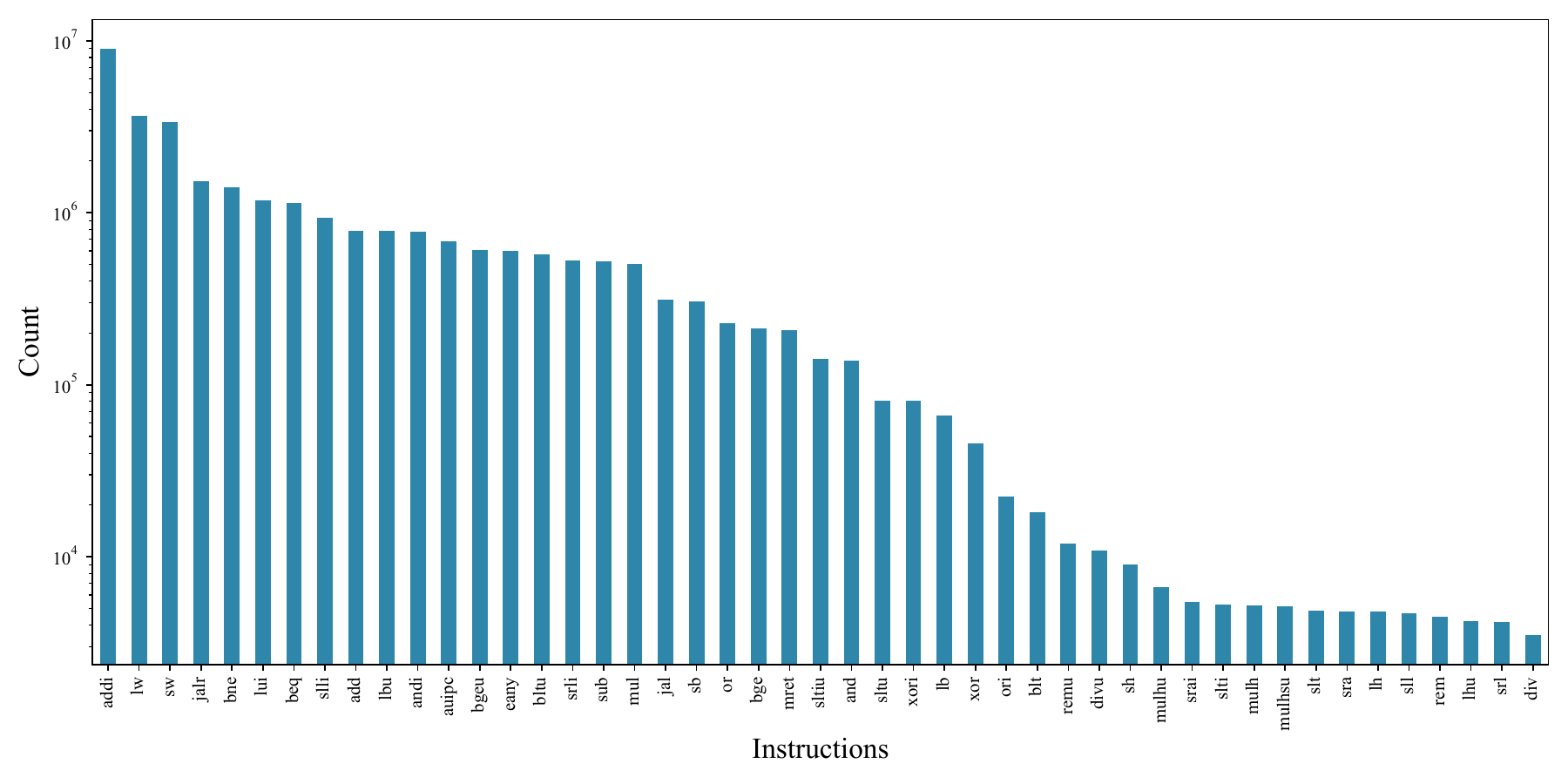}
  \end{subfigure}
  \caption{Instruction-frequency distributions in binaries generated
    from \tool programs for \risc, with (bottom) and without (top) the
    inline-assembly extension. Instructions marked with \code{(*)} are not
    covered.}
  \label{fig:rq4-risc}
\end{figure*}

\begin{figure*}[t]
  \begin{subfigure}[b]{\textwidth}
    \centering
    \includegraphics[scale=0.55]{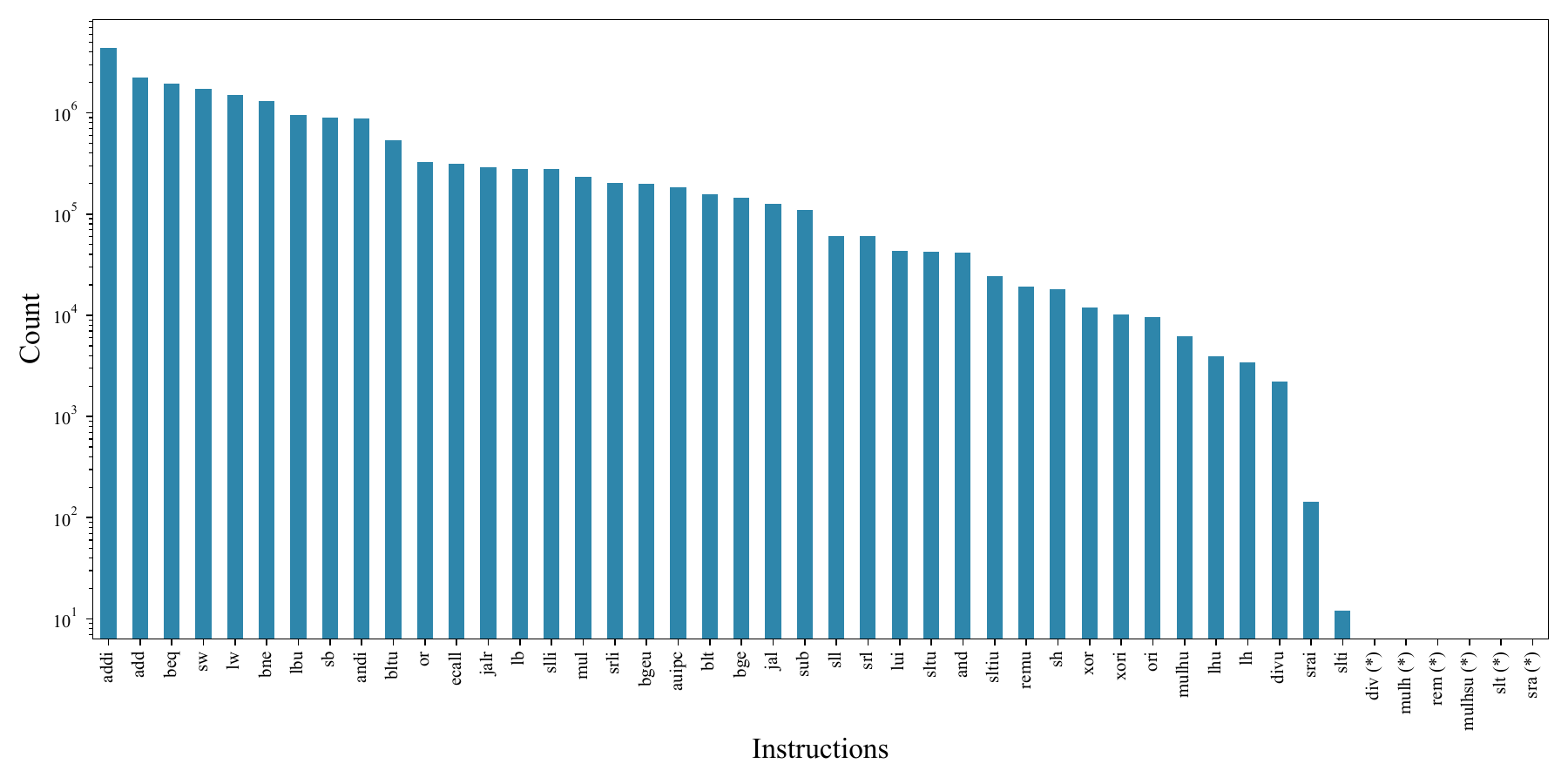}
  \end{subfigure}\\
  \begin{subfigure}[b]{\textwidth}
    \centering
    \includegraphics[scale=0.55]{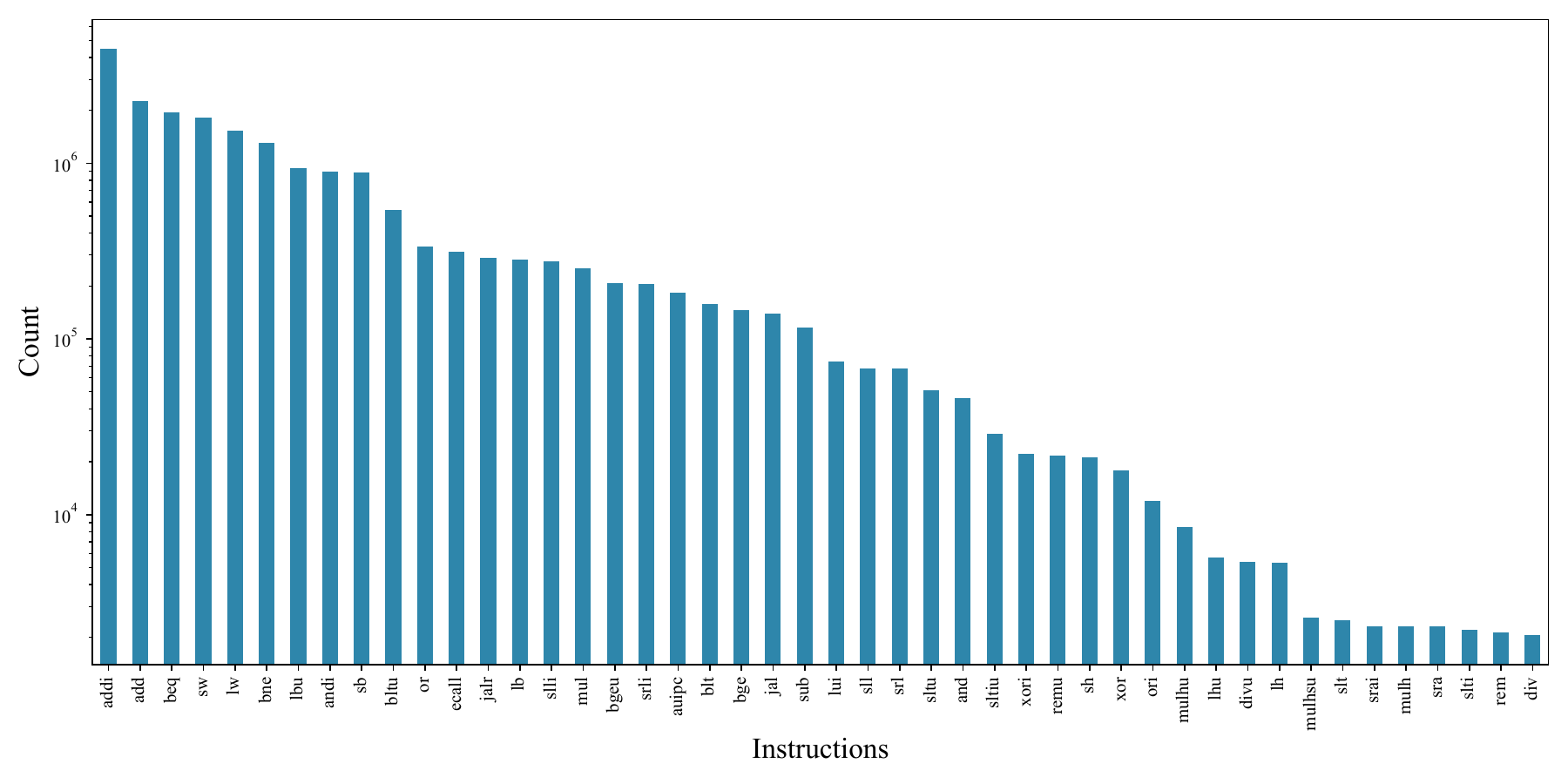}
  \end{subfigure}
  \caption{Instruction-frequency distributions in binaries generated
    from \tool programs for \nexus, with (bottom) and without (top) the
    inline-assembly extension. Instructions marked with \code{(*)} are not
    covered.}
  \label{fig:rq4-nexus}
\end{figure*}

\begin{figure*}[t]
  \begin{subfigure}[b]{\textwidth}
    \centering
    \includegraphics[scale=0.55]{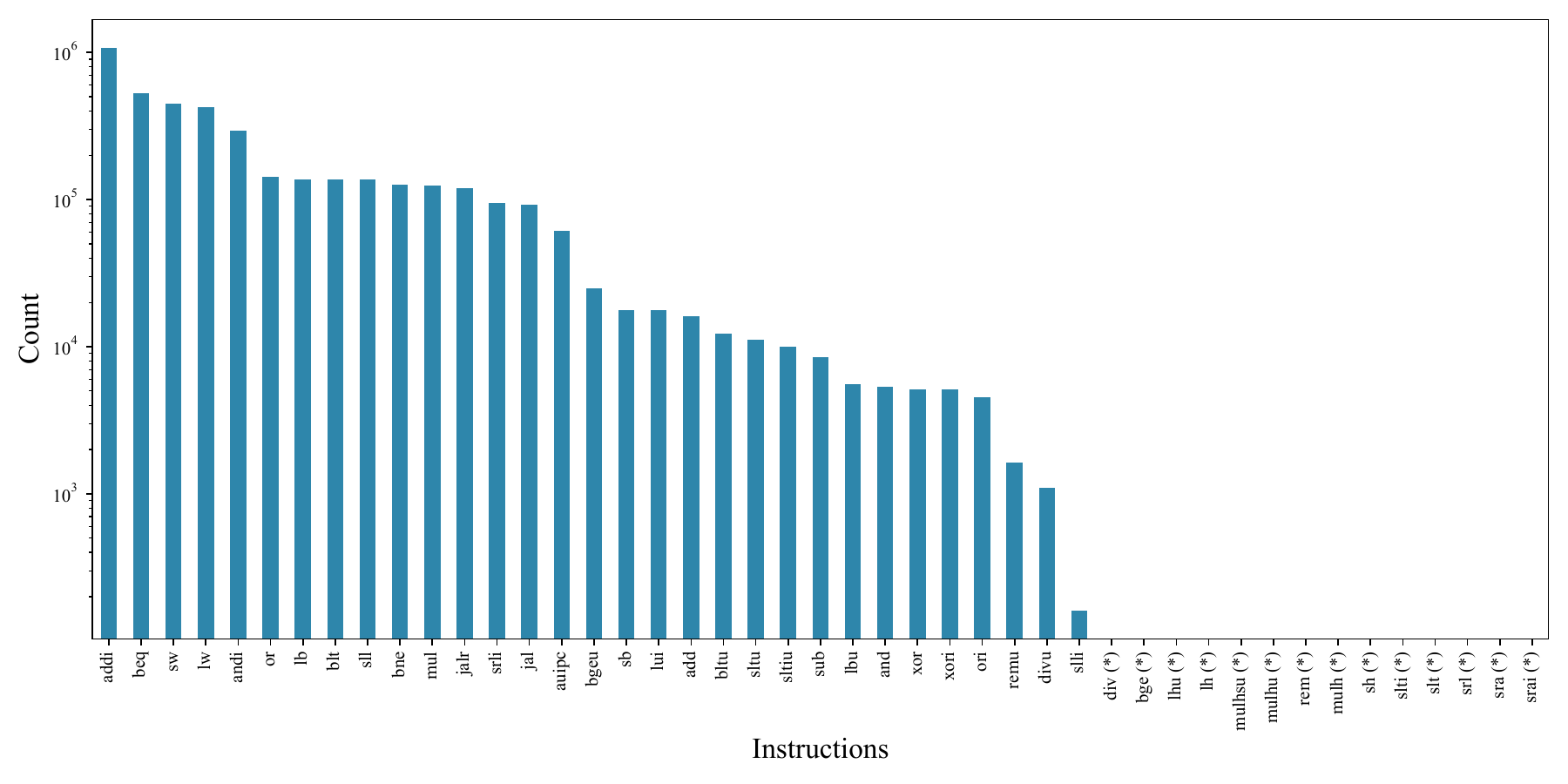}
  \end{subfigure}\\
  \begin{subfigure}[b]{\textwidth}
    \centering
    \includegraphics[scale=0.55]{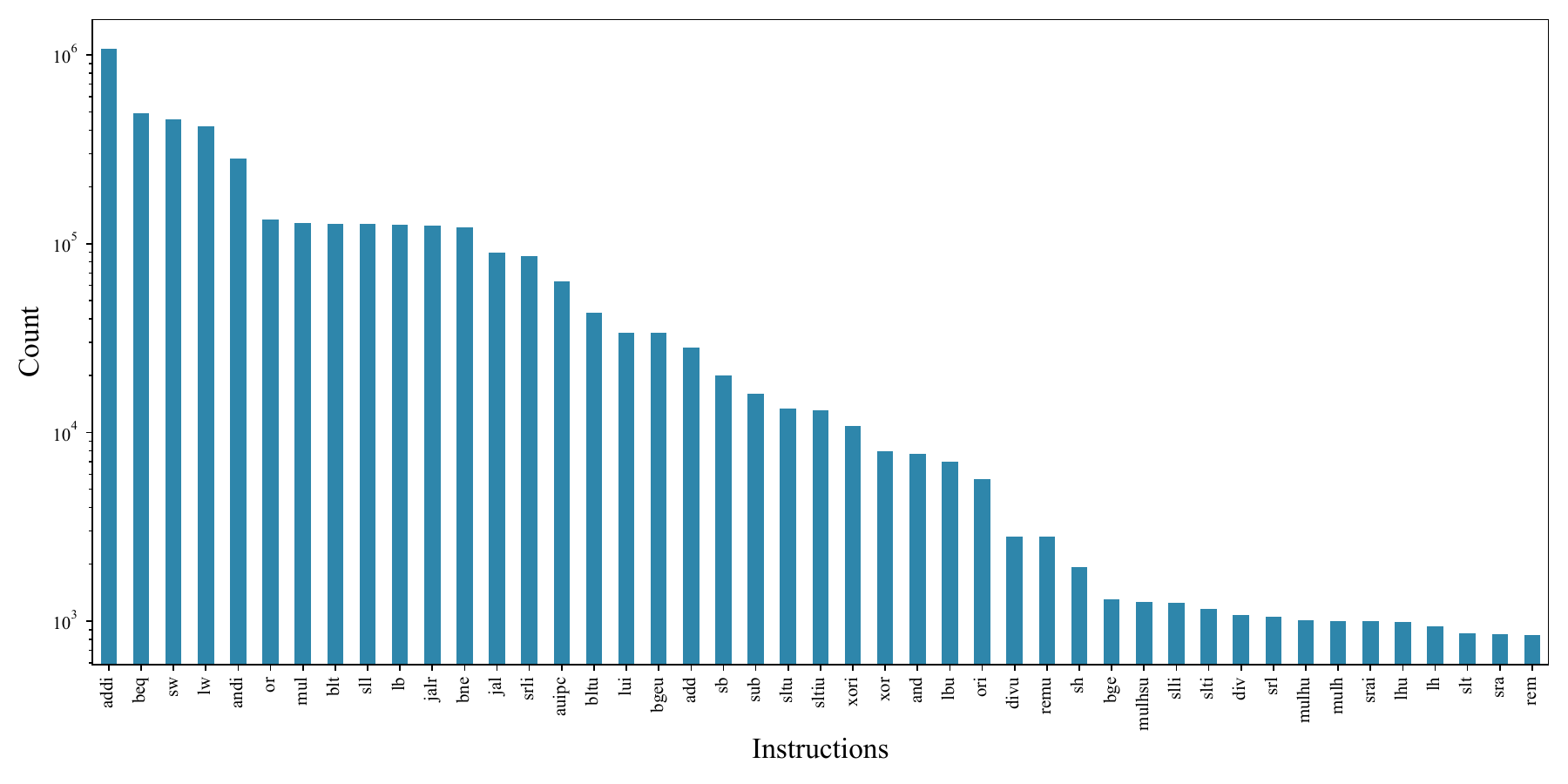}
  \end{subfigure}
  \caption{Instruction-frequency distributions in binaries generated
    from \tool programs for \jolt, with (bottom) and without (top) the
    inline-assembly extension. Instructions marked with \code{(*)} are not
    covered.}
  \label{fig:rq4-jolt}
\end{figure*}

\begin{figure*}[t]
  \begin{subfigure}[b]{\textwidth}
    \centering
    \includegraphics[scale=0.55]{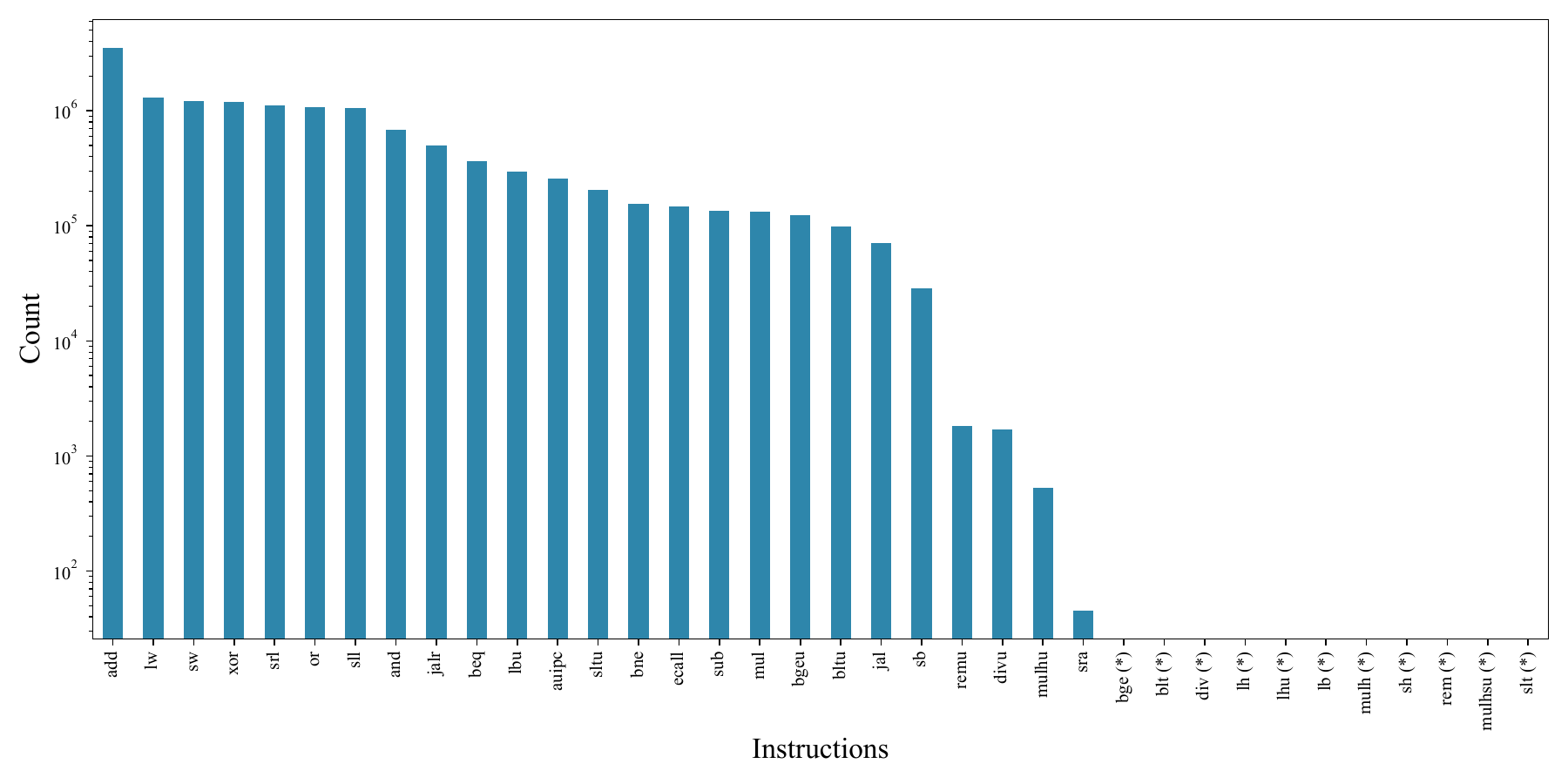}
  \end{subfigure}\\
  \begin{subfigure}[b]{\textwidth}
    \centering
    \includegraphics[scale=0.55]{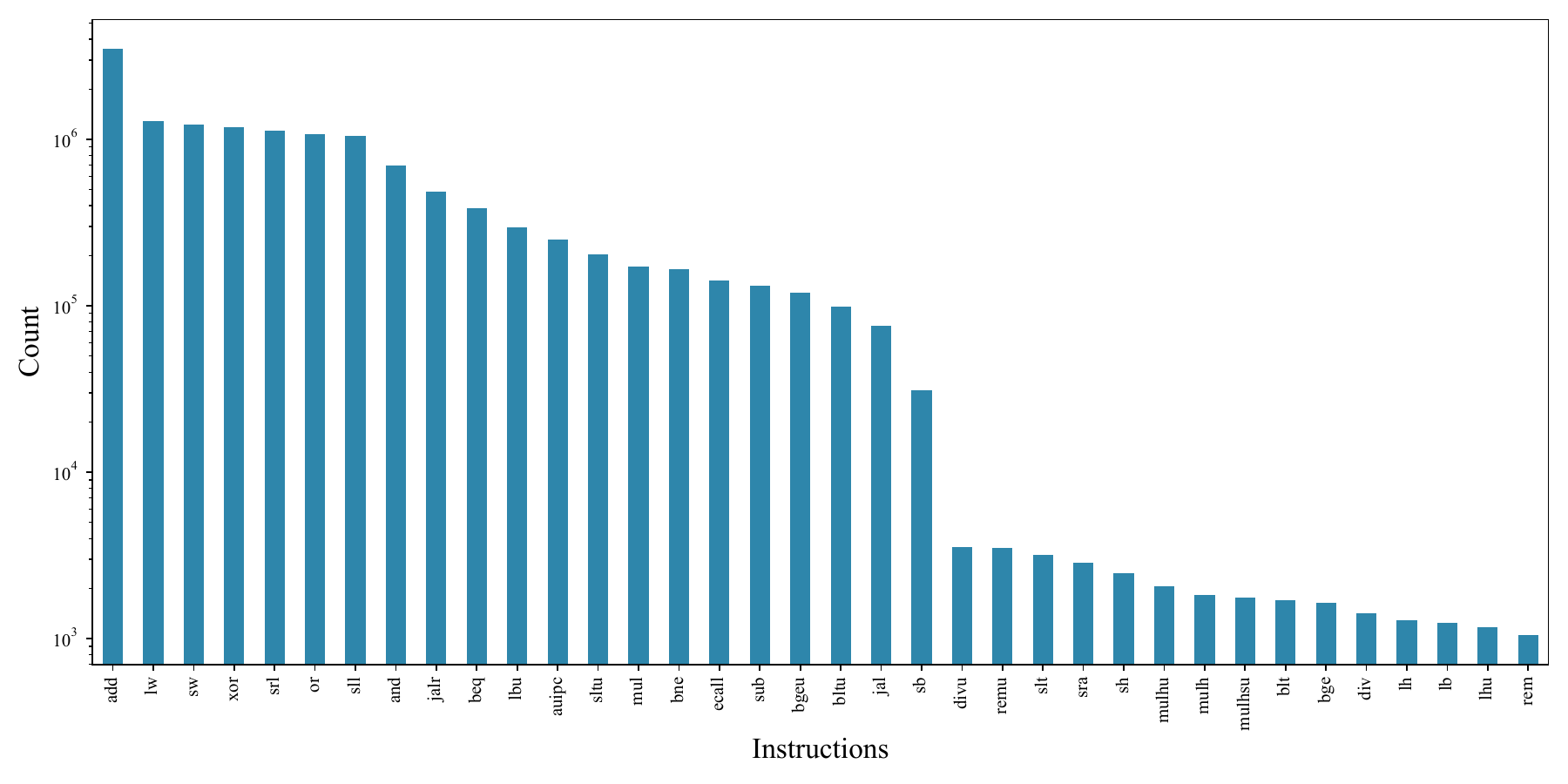}
  \end{subfigure}
  \caption{Instruction-frequency distributions in binaries generated
    from \tool programs for \spOne, with (bottom) and without (top) the
    inline-assembly extension. Instructions marked with \code{(*)} are not
    covered.}
  \label{fig:rq4-sp1}
\end{figure*}

\begin{figure*}[t]
  \begin{subfigure}[b]{\textwidth}
    \centering
    \includegraphics[scale=0.55]{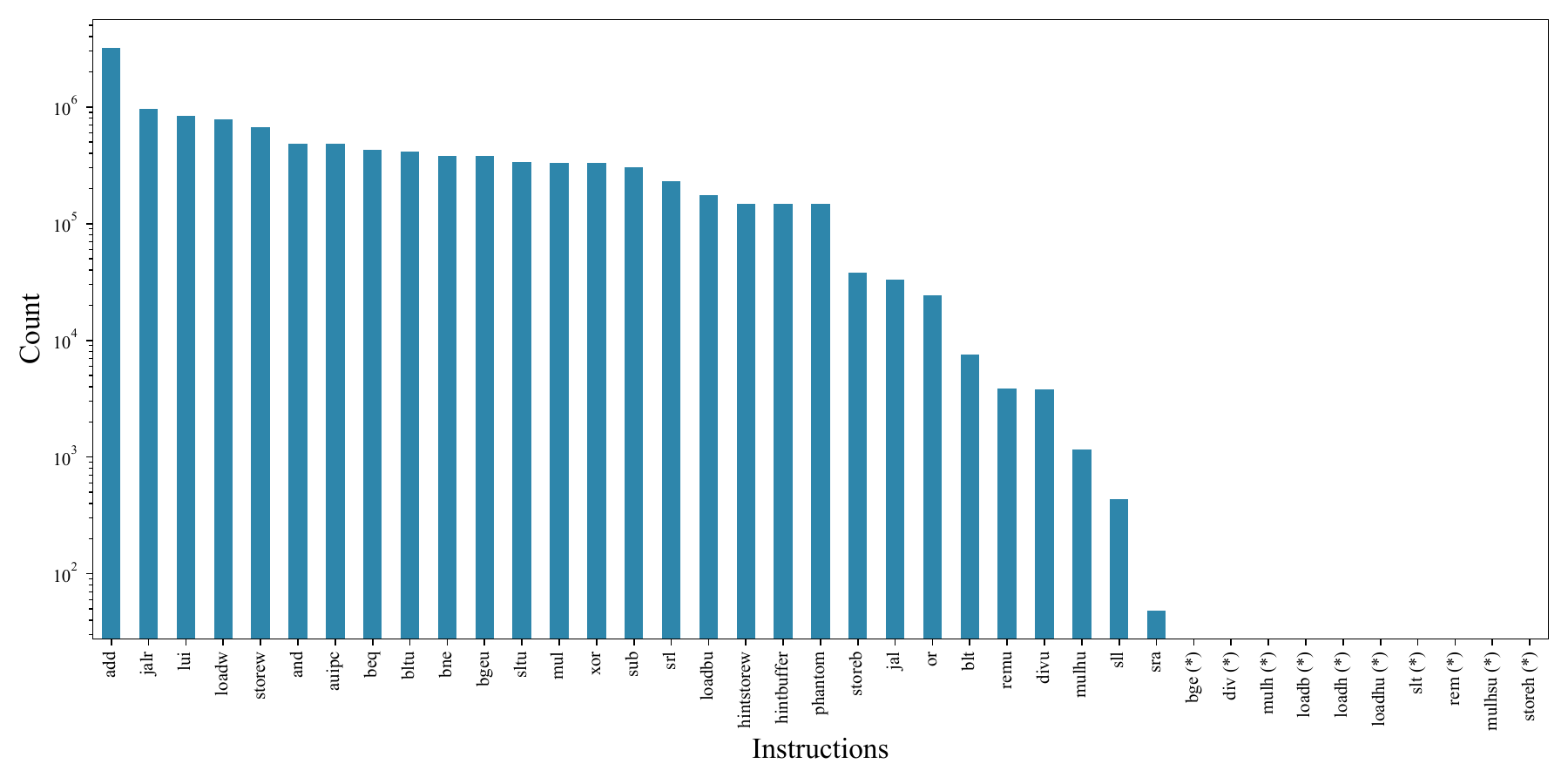}
  \end{subfigure}\\
  \begin{subfigure}[b]{\textwidth}
    \centering
    \includegraphics[scale=0.55]{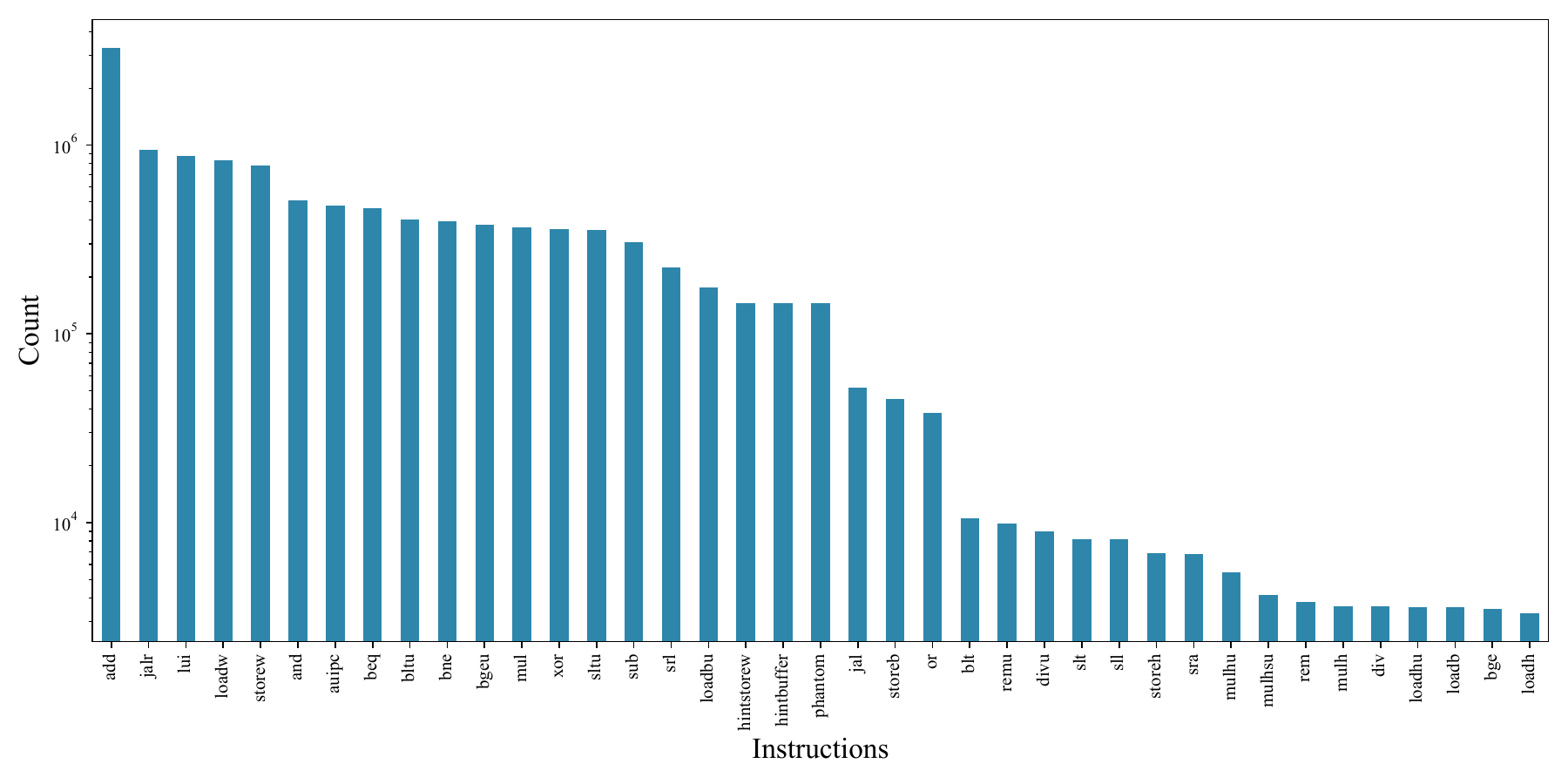}
  \end{subfigure}
  \caption{Instruction-frequency distributions in binaries generated
    from \tool programs for \openvm, with (bottom) and without (top) the
    inline-assembly extension. Instructions marked with \code{(*)} are not
    covered.}
  \label{fig:rq4-openvm}
\end{figure*}

\begin{figure*}[t]
  \begin{subfigure}[b]{\textwidth}
    \centering
    \includegraphics[scale=0.55]{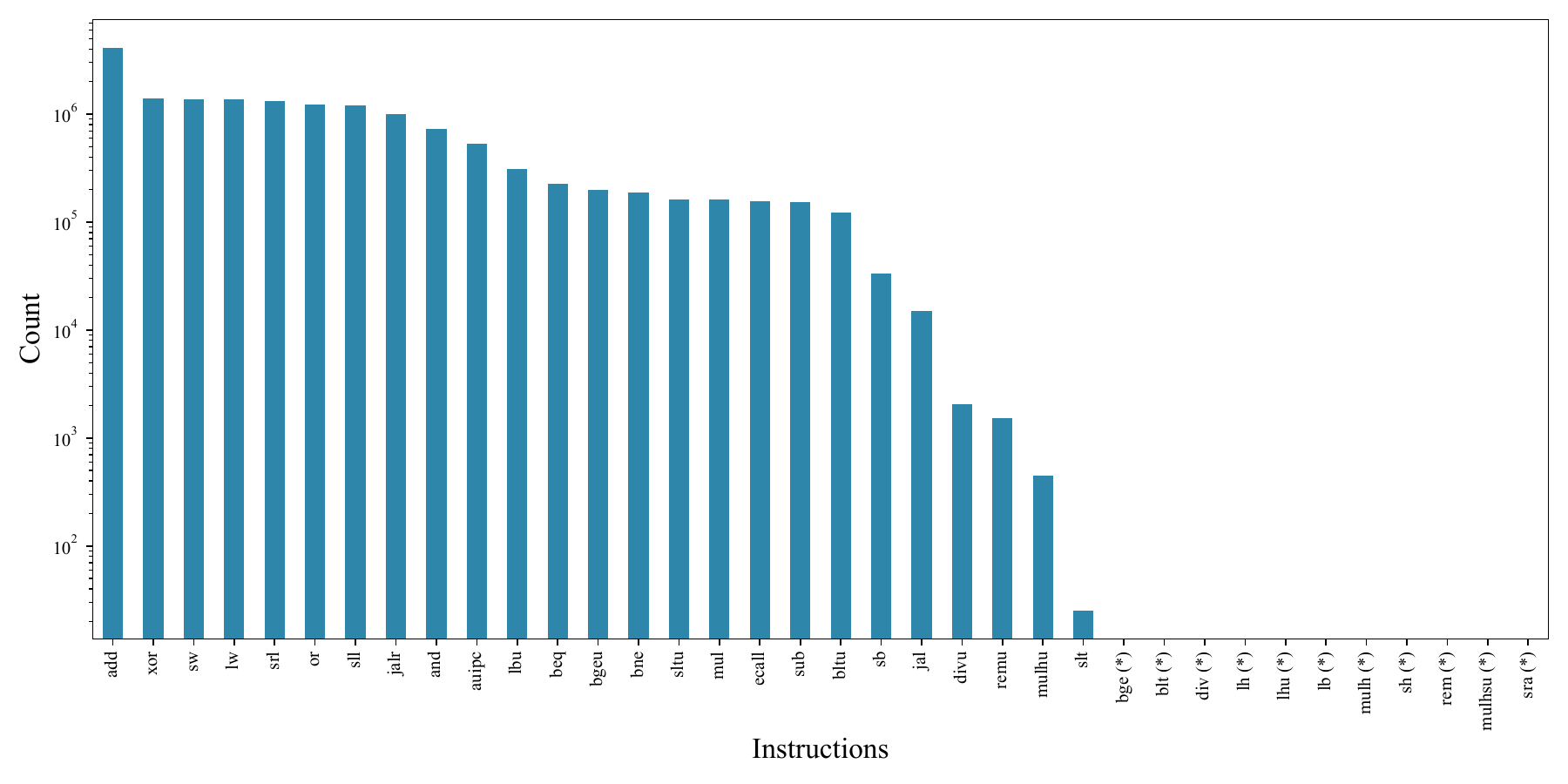}
  \end{subfigure}\\
  \begin{subfigure}[b]{\textwidth}
    \centering
    \includegraphics[scale=0.55]{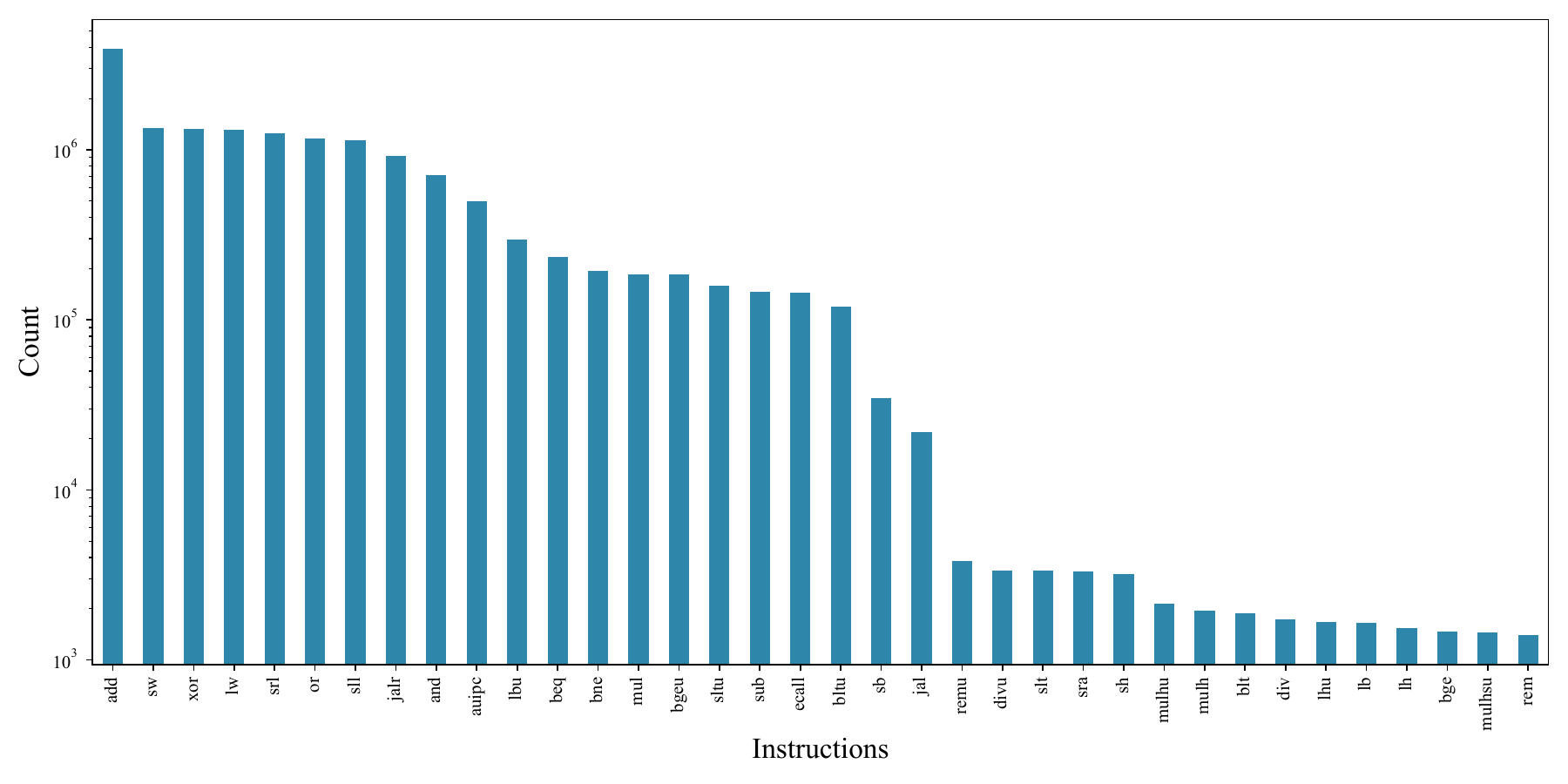}
  \end{subfigure}
  \caption{Instruction-frequency distributions in binaries generated
    from \tool programs for \pico, with (bottom) and without (top) the
    inline-assembly extension. Instructions marked with \code{(*)} are not
    covered.}
  \label{fig:rq4-pico}
\end{figure*}

\figsref{fig:rq4-risc}--\ref{fig:rq4-pico} show, for each zkVM, bar
charts of the instruction-frequency distributions in binaries
generated from \tool programs, with and without the inline-assembly
extension (see RQ4). The x-axis denotes instructions and the y-axis
their frequency; instructions marked with \code{(*)} are not
covered. As the figures illustrate, a significant number of additional
instructions appear in the binaries when the inline-assembly extension
is enabled.

%%% Local Variables:
%%% mode: latex
%%% TeX-master: "main"
%%% End:

\cleardoublepage

\bibliographystyle{plain}
\bibliography{bibliography}

%%%%%%%%%%%%%%%%%%%%%%%%%%%%%%%%%%%%%%%%%%%%%%%%%%%%%%%%%%%%%%%%%%%%%%%%%%%%%%%%
\end{document}